\pdfoutput=1
\documentclass[12pt,a4paper]{article}
\usepackage[colorinlistoftodos]{todonotes}
\usepackage{subfig}
\usepackage{wrapfig}

\usepackage{ifthen} % for conditional statements
\newboolean{pdflatex}
\setboolean{pdflatex}{true} % False for eps figures 

\newboolean{articletitles}
\setboolean{articletitles}{true} % False removes titles in references

\newboolean{uprightparticles}
\setboolean{uprightparticles}{true} %True for upright particle symbols

\newboolean{inbibliography}
\setboolean{inbibliography}{false} %True once you enter the bibliography

\def\paperauthors{Dylan Bourgeois, Conor Fitzpatrick, Sascha Stahl} % Leave as is for PAPER and CONF
\def\paperasciititle{Using holistic event information in the trigger} % Set ASCII title here
\def\papertitle{Using holistic event information in the trigger} % Latex formatted title
\def\paperkeywords{{High Energy Physics}, {Machine Learning}, {Trigger}, {Upgrade}, {LHCb}} % Comma separated list
\def\papercopyright{\the\year\ CERN for the benefit of the LHCb collaboration} % new since 9/Apr/2018
\def\paperlicence{CC-BY-4.0 licence}
\def\paperlicenceurl{https://creativecommons.org/licenses/by/4.0/}

\newcommand{\xhdr}[1]{\vspace{2mm}\noindent{{\bf #1}}}

\usepackage[top=1in, bottom=1.25in, left=1in, right=1in]{geometry}

\usepackage{microtype}
\usepackage{lineno}  % for line numbering during review
\usepackage{xspace} % To avoid problems with missing or double spaces after
\usepackage{caption} %these three command get the figure and table captions automatically small

\usepackage{graphicx}  % to include figures (can also use other packages)
\usepackage{color}
\usepackage{colortbl}
\graphicspath{{./figs/}} % Make Latex search fig subdir for figures
\DeclareGraphicsExtensions{.pdf,.PDF,png,.PNG}

\usepackage{amsmath} % Adds a large collection of math symbols
\usepackage{amssymb}
\usepackage{amsfonts}
\usepackage{upgreek} % Adds in support for greek letters in roman typeset

\newcommand*\patchAmsMathEnvironmentForLineno[1]{%
\expandafter\let\csname old#1\expandafter\endcsname\csname #1\endcsname
\expandafter\let\csname oldend#1\expandafter\endcsname\csname
end#1\endcsname
 \renewenvironment{#1}%
   {\linenomath\csname old#1\endcsname}%
   {\csname oldend#1\endcsname\endlinenomath}%
}
\newcommand*\patchBothAmsMathEnvironmentsForLineno[1]{%
  \patchAmsMathEnvironmentForLineno{#1}%
  \patchAmsMathEnvironmentForLineno{#1*}%
}
\AtBeginDocument{%
\patchBothAmsMathEnvironmentsForLineno{equation}%
\patchBothAmsMathEnvironmentsForLineno{align}%
\patchBothAmsMathEnvironmentsForLineno{flalign}%
\patchBothAmsMathEnvironmentsForLineno{alignat}%
\patchBothAmsMathEnvironmentsForLineno{gather}%
\patchBothAmsMathEnvironmentsForLineno{multline}%
\patchBothAmsMathEnvironmentsForLineno{eqnarray}%
}

\usepackage{hyperxmp}

\usepackage[pdftex,
            pdfauthor={\paperauthors},
            pdftitle={\paperasciititle},
            pdfkeywords={\paperkeywords},
            pdfcopyright={Copyright (C) \papercopyright},
            pdflicenseurl={\paperlicenceurl}]{hyperref}

\usepackage[all]{hypcap} % Internal hyperlinks to floats.

\usepackage{xspace} 
\usepackage{upgreek}

\def\lhcb   {\mbox{LHCb}\xspace}

\def\lhc    {\mbox{LHC}\xspace}

\def\velo   {VELO\xspace}
\def\rich   {RICH\xspace}

\def\ecal   {ECAL\xspace}
\def\hcal   {HCAL\xspace}
\def\MagUp {\mbox{\em Mag\kern -0.05em Up}\xspace}

\ifthenelse{\boolean{uprightparticles}}%
{

 \def\PDelta      {\ensuremath{\Delta}\xspace}                 
 \def\PXi      {\ensuremath{\Xi}\xspace}                 
 \def\PLambda      {\ensuremath{\Lambda}\xspace}                 
 \def\PSigma      {\ensuremath{\Sigma}\xspace}                 
 \def\POmega      {\ensuremath{\Omega}\xspace}                 
 \def\PUpsilon      {\ensuremath{\Upsilon}\xspace}                 
 
 \def\PB      {\ensuremath{\mathrm{B}}\xspace}                 
                  
 \def\PD      {\ensuremath{\mathrm{D}}\xspace}

 \def\PK      {\ensuremath{\mathrm{K}}\xspace}

 \def\Pi      {\ensuremath{\mathrm{i}}\xspace}

}
{

 \mathchardef\PDelta="7101
 \mathchardef\PXi="7104
 \mathchardef\PLambda="7103
 \mathchardef\PSigma="7106
 \mathchardef\POmega="710A
 \mathchardef\PUpsilon="7107
                  
 \def\PB      {\ensuremath{B}\xspace}                 
                  
 \def\PD      {\ensuremath{D}\xspace}

 \def\PK      {\ensuremath{K}\xspace}

 \def\Pi      {\ensuremath{i}\xspace}

}

\makeatletter
\ifcase \@ptsize \relax% 10pt
  \newcommand{\miniscule}{\@setfontsize\miniscule{4}{5}}% \tiny: 5/6
\or% 11pt
  \newcommand{\miniscule}{\@setfontsize\miniscule{5}{6}}% \tiny: 6/7
\or% 12pt
  \newcommand{\miniscule}{\@setfontsize\miniscule{5}{6}}% \tiny: 6/7
\fi
\makeatother

\DeclareRobustCommand{\optbar}[1]{\shortstack{{\miniscule (\rule[.5ex]{1.25em}{.18mm})}
  \\ [-.7ex] $#1$}}

  \def\Kbar    {{\kern 0.2em\overline{\kern -0.2em \PK}{}}\xspace}

\def\KorKbar    {\kern 0.18em\optbar{\kern -0.18em K}{}\xspace}

  \def\Dbar    {{\kern 0.2em\overline{\kern -0.2em \PD}{}}\xspace}

\def\DorDbar    {\kern 0.18em\optbar{\kern -0.18em D}{}\xspace}

\def\B       {{\ensuremath{\PB}}\xspace}
\def\Bbar    {{\ensuremath{\kern 0.18em\overline{\kern -0.18em \PB}{}}}\xspace}

\def\BorBbar    {\kern 0.18em\optbar{\kern -0.18em B}{}\xspace}

  \def\Y#1S{\ensuremath{\PUpsilon{(#1S)}}\xspace}% no space before {...}!

\def\Lbar        {{\ensuremath{\kern 0.1em\overline{\kern -0.1em\PLambda}}}\xspace}
\def\LorLbar    {\kern 0.18em\optbar{\kern -0.18em \PLambda}{}\xspace}

\def\AT#1     {\ensuremath{A_{\mathrm{T}}^{#1}}\xspace}           % 2

\def\C#1      {\ensuremath{\mathcal{C}_{#1}}\xspace}                       % 9
\def\Cp#1     {\ensuremath{\mathcal{C}_{#1}^{'}}\xspace}                    % 7
\def\Ceff#1   {\ensuremath{\mathcal{C}_{#1}^{\mathrm{(eff)}}}\xspace}        % 9  
\def\Cpeff#1  {\ensuremath{\mathcal{C}_{#1}^{'\mathrm{(eff)}}}\xspace}       % 7
\def\Ope#1    {\ensuremath{\mathcal{O}_{#1}}\xspace}                       % 2
\def\Opep#1   {\ensuremath{\mathcal{O}_{#1}^{'}}\xspace}                    % 7

\newcommand{\tev}{\ifthenelse{\boolean{inbibliography}}{\ensuremath{~T\kern -0.05em eV}}{\ensuremath{\mathrm{\,Te\kern -0.1em V}}}\xspace}
\newcommand{\gev}{\ensuremath{\mathrm{\,Ge\kern -0.1em V}}\xspace}
\newcommand{\mev}{\ensuremath{\mathrm{\,Me\kern -0.1em V}}\xspace}
\newcommand{\kev}{\ensuremath{\mathrm{\,ke\kern -0.1em V}}\xspace}
\newcommand{\ev}{\ensuremath{\mathrm{\,e\kern -0.1em V}}\xspace}
\newcommand{\gevc}{\ensuremath{{\mathrm{\,Ge\kern -0.1em V\!/}c}}\xspace}
\newcommand{\mevc}{\ensuremath{{\mathrm{\,Me\kern -0.1em V\!/}c}}\xspace}
\newcommand{\gevcc}{\ensuremath{{\mathrm{\,Ge\kern -0.1em V\!/}c^2}}\xspace}
\newcommand{\gevgevcccc}{\ensuremath{{\mathrm{\,Ge\kern -0.1em V^2\!/}c^4}}\xspace}
\newcommand{\mevcc}{\ensuremath{{\mathrm{\,Me\kern -0.1em V\!/}c^2}}\xspace}

\def\mhz  {\ensuremath{{\mathrm{ \,MHz}}}\xspace}

\def\gsim{{~\raise.15em\hbox{$>$}\kern-.85em
          \lower.35em\hbox{$\sim$}~}\xspace}
\def\lsim{{~\raise.15em\hbox{$<$}\kern-.85em
          \lower.35em\hbox{$\sim$}~}\xspace}

\def\davinci    {\mbox{\textsc{DaVinci}}\xspace}
\def\dirac      {\mbox{\textsc{Dirac}}\xspace}

\def\root       {\mbox{\textsc{Root}}\xspace}

\def\tell1  {TELL1\xspace}
\def\ukl1   {UKL1\xspace}

\usepackage{cite} % Allows for ranges in citations
\usepackage{mciteplus}

\usepackage{longtable} % only for template; not usually to be used in PAPERs

\begin{document}

%%%%%%%%%%%%%%%%%%%%%%%%%
%%%%% Title     %%%%%%%%%
%%%%%%%%%%%%%%%%%%%%%%%%%
\renewcommand{\thefootnote}{\fnsymbol{footnote}}
\setcounter{footnote}{1}

% %%%%%%% CHOOSE TITLE PAGE--------
%\onecolumn
%\input{title-LHCb-INT}
%\input{title-LHCb-ANA}
%\input{title-LHCb-CONF}
% $Id: title-LHCb-PAPER.tex 120776 2018-05-29 05:11:17Z pkoppenb $
% ===============================================================================
% Purpose: LHCb-PAPER journal paper title page template
% Author:
% Created on: 2010-09-25
% ===============================================================================

%%%%%%%%%%%%%%%%%%%%%%%%%
%%%%%  TITLE PAGE  %%%%%%
%%%%%%%%%%%%%%%%%%%%%%%%%
\begin{titlepage}
\pagenumbering{roman}

% Header ---------------------------------------------------
\vspace*{-1.5cm}
\centerline{\large EUROPEAN ORGANIZATION FOR NUCLEAR RESEARCH (CERN)}
\vspace*{1.5cm}
\noindent
\begin{tabular*}{\linewidth}{lc@{\extracolsep{\fill}}r@{\extracolsep{0pt}}}
\ifthenelse{\boolean{pdflatex}}% Logo format choice
{\vspace*{-1.2cm}\mbox{\!\!\!\includegraphics[width=.14\textwidth]{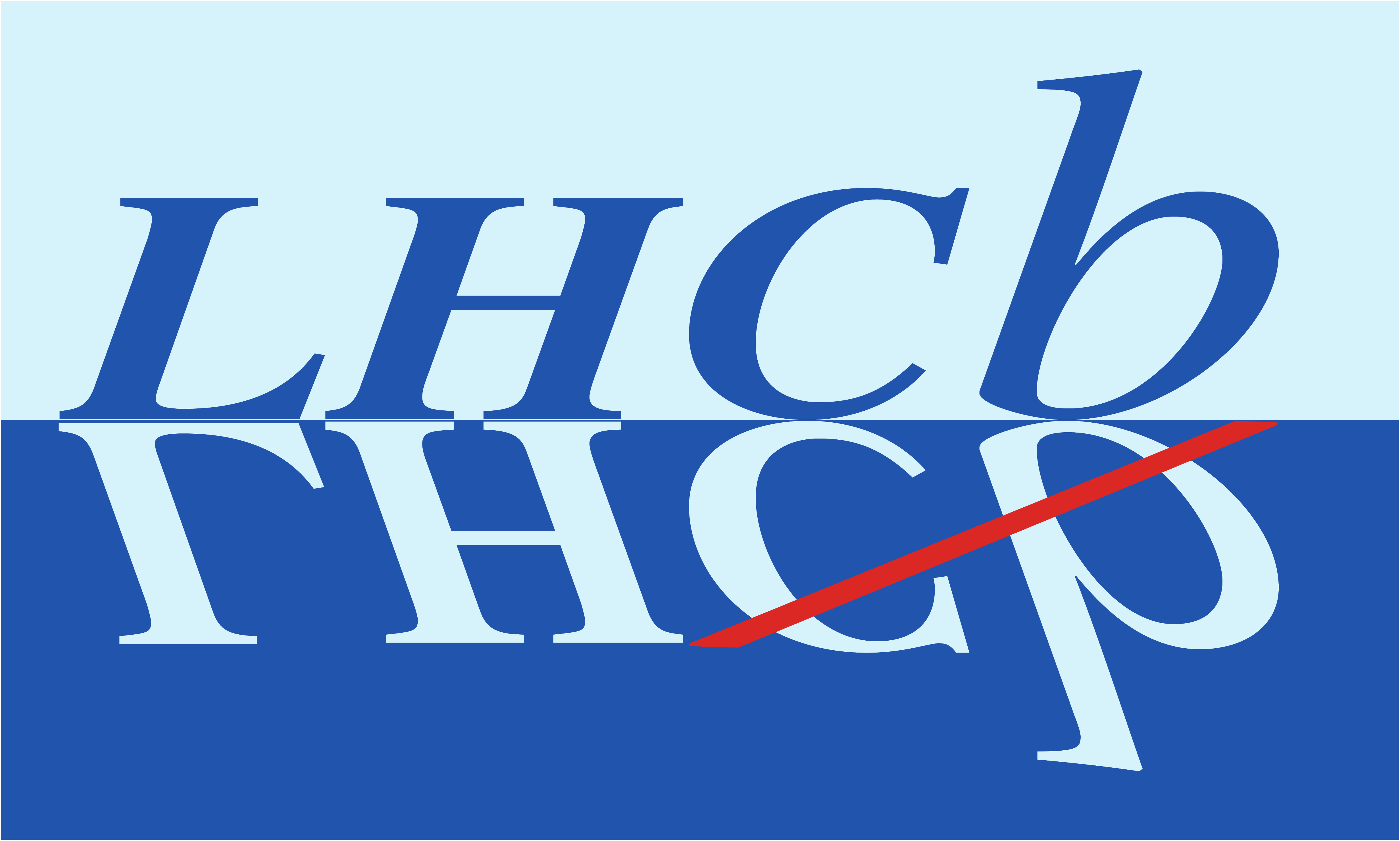}} & &}%
{\vspace*{-1.2cm}\mbox{\!\!\!\includegraphics[width=.12\textwidth]{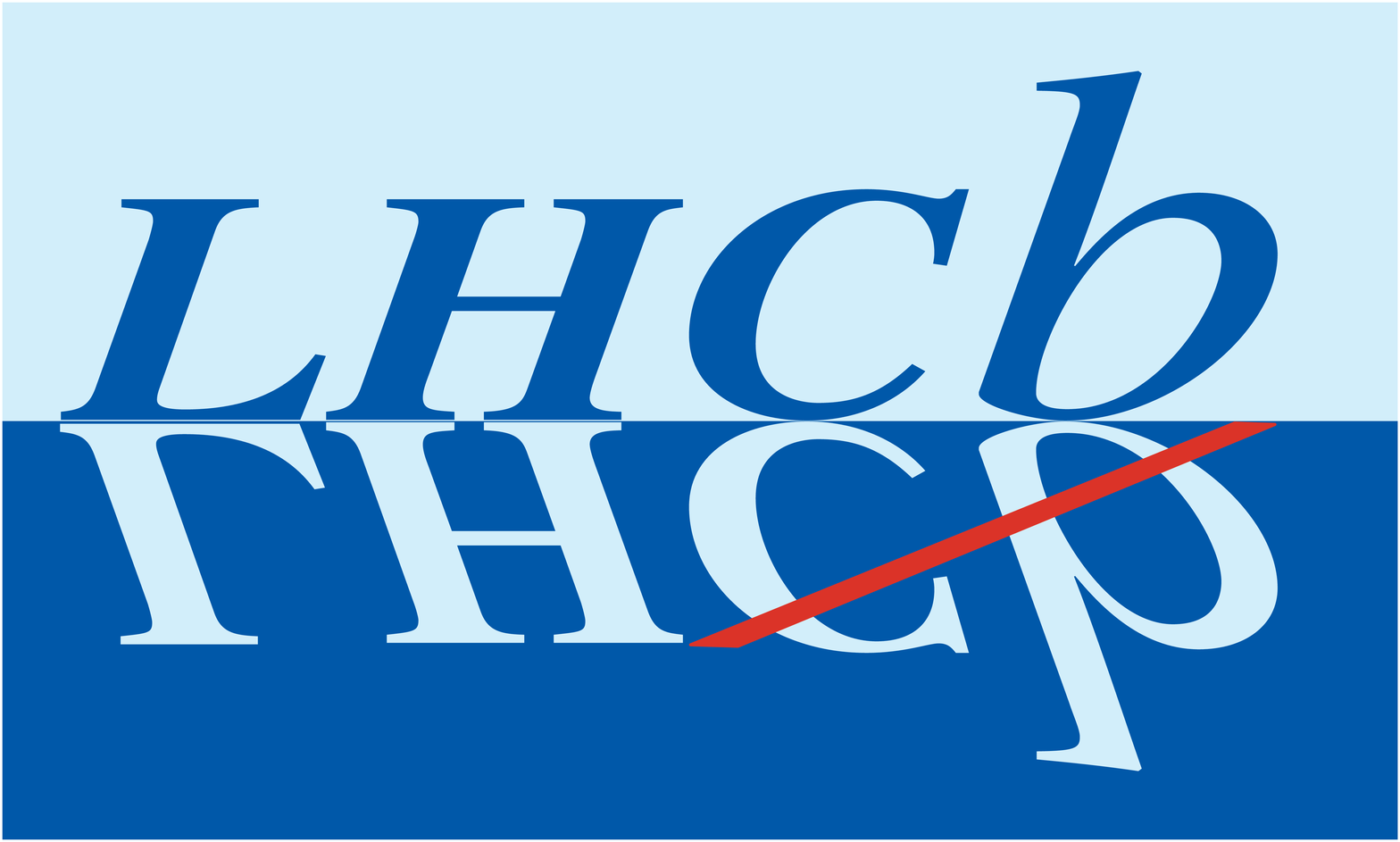}} & &}
\\
 %& & CERN-EP-2018-010 \\  % ID 
 & & LHCb-PUB-2018-010 \\  % ID
 & & \today \\ % Date - Can also hardwire e.g.: 23 March 2010
 & & \\
% not in paper \hline
\end{tabular*}

\vspace*{4.0cm}

% Title --------------------------------------------------
{\normalfont\bfseries\boldmath\huge
\begin{center}
% DO NOT EDIT HERE. Instead edit macro in main.tex to keep metadata correct
  \papertitle
\end{center}
}

\vspace*{2.0cm}

% Authors -------------------------------------------------
\begin{center}
% If changing to list here, make pdfauthors in main.tex a comma
% separated list with the same names. Otherwise metadata in file will be wrong.
Dylan Bourgeois$^1$\footnote{Corresponding author}, Conor Fitzpatrick$^{1}$, Sascha Stahl$^2$.
\bigskip\\
{\normalfont\itshape\footnotesize
$ ^1$\'{E}cole Polytechnique F\'{e}d\'{e}rale de Lausanne, Switzerland\\
$ ^2$CERN\\
}
\end{center}

\vspace{\fill}

% Abstract -----------------------------------------------
\begin{abstract}
  \noindent

  In order to achieve the data rates proposed for the future Run 3 upgrade of the \lhcb detector, new processing models must be developed to deal with the increased throughput. For this reason we aim to investigate the feasibility of purely data-driven ``holistic'' methods, with the constraint of introducing minimal computational overhead, hence using only raw detector information. These filters should be unbiased - having a neutral effect with respect to the studied physics channels. In particular, the use of machine learning based methods seems particularly suitable, potentially providing a natural formulation for heuristic-free, unbiased filters whose objective would be to optimize between throughput and bandwidth.

\end{abstract}

\vspace*{2.0cm}

%\begin{center}
%  Submitted to JHEP / Phys.~Rev.~D / Phys.~Rev.~Lett. / Phys.~Lett.~B / Eur.~Phys.~J.~C / Nucl.~Phys.~B
%\end{center}

\vspace{\fill}

{\footnotesize
% Edit macro in main.tex to keep metadata correct
\centerline{\copyright~\papercopyright. \href{\paperlicenceurl}{\paperlicence}.}}
\vspace*{2mm}

\end{titlepage}

%%%%%%%%%%%%%%%%%%%%%%%%%%%%%%%%
%%%%%  EOD OF TITLE PAGE  %%%%%%
%%%%%%%%%%%%%%%%%%%%%%%%%%%%%%%%

%  empty page follows the title page ----
\newpage
\setcounter{page}{2}
\mbox{~}
%\newpage
%
%% Author List ----------------------------
%%  You need to get a new author list!
%\input{LHCb_authorlist.tex}
%
%The author list for journal publications is provided by the Membership Committee shortly after 'approval to go to paper' has been given.
%%It will be made available on the page
%%\verb!http://www.physik.uzh.ch/~strauman/forMemCo/LHCb-PAPER-XXXX-XXX/! .
%It will be sent to you by email shortly after a paper number has beens assigned.
%The author list should be included already at first circulation,
%to allow new members of the collaboration to verify whether they have been included correctly.
%Occasionally a misspelled name is corrected or associated institutions become full members.
%In that case, a new author list will be sent to you.
%In case line numbering doesn't work well after including the authorlist, try moving the \verb!\bigskip! after the last author to a separate line.
%
%
%The authorship for Conference Reports should be ``The LHCb
%  collaboration'', with a footnote giving the name(s) of the contact
%  author(s), but without the full list of collaboration names.

\cleardoublepage

%\twocolumn
% %%%%%%%%%%%%% ---------

\renewcommand{\thefootnote}{\arabic{footnote}}
\setcounter{footnote}{0}

%%%%%%%%%%%%%%%%%%%%%%%%%%%%%%%%
%%%%%  Table of Content   %%%%%%
%%%%%%%%%%%%%%%%%%%%%%%%%%%%%%%%
%%%% Uncomment next 2 lines if desired
%\tableofcontents
%\newpage
% \cleardoublepage

%%%%%%%%%%%%%%%%%%%%%%%%%
%%%%% Main text %%%%%%%%%
%%%%%%%%%%%%%%%%%%%%%%%%%

\pagestyle{plain} % restore page numbers for the main text
\setcounter{page}{1}
\pagenumbering{arabic}

%% Uncomment during review phase. 
%% Comment before a final submission.
%\linenumbers

\section{Introduction}
\label{sec:Introduction}

The increase in instantaneous luminosity delivered by the \lhc in Run 3 will lead to a rise in the event rate to be processed by the \lhcb experiment. As detailed in the \lhcb Trigger and Online Upgrade TDR~\cite{LHCb-TDR-012}, there is a strong case to be made that higher trigger output rates would allow to fully exploit both beauty and charm physics programs. This means the trigger will have to deal with more data coming in but also output more data than ever before. A redesign of the trigger architecture is therefore required to deal with these high data rates. Additionally, the desire to calibrate and align all subdetectors online - with data collected in real-time by the High-Level Trigger (HLT) - requires that the increase in event processing power be made available for real-time settings, not just in the offline farm.

The first parts of the new event processing model were implemented during the first Long Shutdown (LS1) in preparation for Run 2. Currently, the event processing model is structured in three stages: a first hardware level trigger (L0) reduces the rate of the sensor readouts from $30$ \mhz to $1$ \mhz by using information from the calorimeter and muon systems. The selected data is then sent to the Event Filter Farm (EFF), where a two-stage software filter, the High-Level Trigger (HLT), reduces the rate further~\cite{HLTR2}. The first stage (HLT1) is run \textit{during} data taking, performing a partial reconstruction and filtering. When enough resources are available (during data taking if possible, otherwise outside of LHC providing collisions), the second stage (HLT2) performs the complete reconstruction and filtering.

In Run 3 however, the goal is to do away with the hardware trigger (L0), meaning the software-based HLT must deal with rates up to $30$ \mhz, the inelastic proton-proton collision rate given by the \lhc. The obvious first step is to increase the size of the EFF to help cope with the increased rates. An estimate of the cost for an EFF hardware upgrade which would allow it to process this amount of data was made in 2014~\cite{LHCb-TDR-016}. Since then the growth rate of CPU performance has declined~\cite{Aaij:2244312}, meaning that keeping up with these event rates would require further performance improvements to the core software framework, the event data model, and in the implementation of the trigger algorithms~\cite{TheLHCbCollaboration:2310827, DeCian:2309972}. 

This situation opens the door to the investigation of novel methods, notably data-driven solutions. In recent years machine learning has found success in a wide variety of applications. The ability of these algorithms to automatically learn representative and discriminative features that can replace manually engineered ones has proven very useful and effective. Additionally, the newfound availability of large amounts of annotated data has allowed these algorithms to be deployed at scale. Luckily, high-energy physics is one domain in which data is abundant: Monte-Carlo samples serve as a clean source to train these data-hungry algorithms. Several data-centric approaches to the field have been successful, notably under the impulse of the High Energy Physics-Machine Learning (HEP-ML) community~\cite{hepml, Albertsson:2018maf, Guest:2018yhq}.

In the ideal scenario, an end-to-end machine learning pipeline would be developed, with as an objective function the maximization of the available rate with as little loss as possible in the useful event rate. Unfortunately, this objective is clearly not differentiable and is not even explicit. The definition of ``usefulness'' is vague at best, resulting in a concern that any efficiency gain must not come at the cost of parts of the physics program. This motivates the search for an \textit{unbiased} filter, in the sense that it is agnostic to the desired output but only partial to the input distribution. Data-driven approaches can be a double-edged sword in this regard since they usually incorporate little to no prior knowledge~\cite{Brehmer:2018hga, Cranmer:2015bka}. The absence of heuristics could, in theory, make the algorithms blind to any domain-specific artifacts. On the other hand, data-driven algorithms are also very sensitive to biases inherent to the distribution of input data. For example, they often exhibit a bias towards the mean, which could penalize lower throughput channels compared to higher throughput (but potentially less important) channels. These behaviors are inherent to the algorithms and are in this regard difficult to manipulate upstream. For this reason, one must take additional care to verify the results a posteriori, beyond simple accuracy measures~\cite{accuracy} and across a wide set of scenarios that could be impacted.\\ \newline

Given all of these considerations, we can formulate a set of research questions that we will investigate in the following work.

\xhdr{Research Questions}

\begin{description}
\item \textbf{RQ1:} What is an optimal trade-off when cutting signal between the computational cost of reconstruction and information content for a given event?
\item \textbf{RQ2:} Are data-driven approaches adapted to learn mappings between the raw detector readouts and useful physical quantities?
\item \textbf{RQ3:} Would filtering on raw detector readouts be unbiased?

\end{description}

In Section~\ref{sec:holistic}, we describe the overall pipeline designed to investigate these questions, including the input data at our disposal as well as possible targets, and present ways to bridge the gap between the two by learning a mapping from input to target. In Section~\ref{sec:experiments}, we present a set of experiments that attempt to answer the questions posed above. In Section~\ref{sec:discussion}, we discuss our general observations and some limitations of our approach, showing opportunities for future work. Finally, we wrap up and present our conclusions in Section~\ref{sec:conclusion}.

\newpage
% ----- SECTION
\section{Using Holistic Information in the Trigger}
\label{sec:holistic}

In this project, we direct our efforts at finding ways to reduce the rate of events that are of no use to the physics program (i.e. background) while introducing a minimal amount of overhead. Consequently, it makes sense to use information that is available before any type of reconstruction. The information we access is detailed in this section.

% ---------------
% ----- 2.1 -----
\subsection{Available readouts}
\label{subsec:input}

When a sensor readout is performed, i.e. when an event occurs, the Data AcQuisition (DAQ) system will package the sensor activity into data-agnostic structures. These are passed through directly to the trigger for initial processing. These raw readouts contain a tiny header (negligible since it is of constant size) and a raw bank of information. By measuring the size (in bytes) of these raw banks, we can have a rough estimate of the activity induced by a given event. This is denoted as ``raw event information'' (sometimes replaced by ``raw bank size'' in the following analyses), and forms the basic data source for our analysis.

\begin{figure}[ht]
  \begin{center}
    \includegraphics[width=1.0\linewidth]{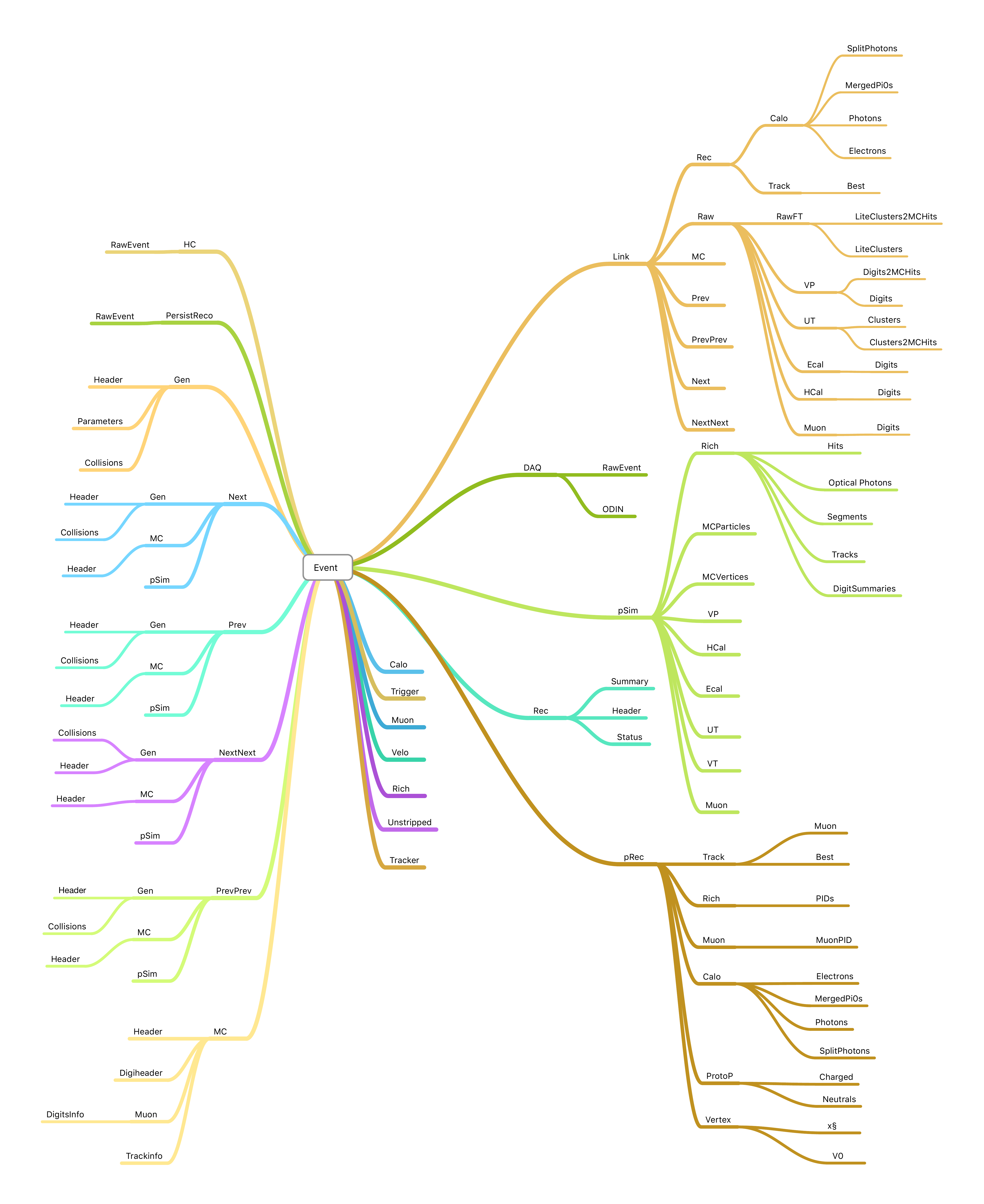}
    \vspace*{-0.5cm}
  \end{center}
  \caption{
  	Tree structure of the different TES locations available for an event in \lhcb event processing model
    }
  \label{fig:tes_tree}
\end{figure}

Specifically, the information is archived in Transient Event Stores (TES), which contain particular locations for each type of raw bank. In our case we want access to the raw event information, which can be found at the \texttt{Event/DAQ/RawEvent} location (see Fig. \ref{fig:tes_tree}). A custom \davinci script~\footnote{https://gitlab.cern.ch/lhcb-HLT/holistic-information-trigger/blob/master/rawEventSize.py} was created to process and extract the required information, with everything from downloading the \texttt{.ldst} files from \dirac to creating custom \root files, which are then converted to \texttt{NumPy} arrays.

% ----- 2.1.1 -----
\subsubsection{Bank Sizes}
\label{ssubsec:rawbank}

At the \texttt{RawEvent} location, there exists a set of raw banks which hold information about the event. We will use the size of the raw bank as a proxy for activity: the more hits a detector registered the larger the raw  bank that contains this information. This is a common approach to estimate the bandwidth of a stripping line for example, as in the \texttt{lhcb-HLT/upgrade-bandwidth-studies}~\footnote{https://gitlab.cern.ch/lhcb-HLT/upgrade-bandwidth-studies} work, which aims to estimate the throughput of selected trigger lines.

\begin{table}
	\centering
\begin{tabular}{ |l|l| }
  \hline
  FullEventSize & Full event bank \\
  \rich 		& Ring-Imaging Cherenkov detector \\
  Muon 			& Muon detector \\
  \ecal 		& Electromagnetic Calorimeter \\
  \hcal 		& Hadron Calorimeter \\
  VP 			& VErtex LOcator (VELO) Pixel detector \\
  FTCluster 	& Forward Tracking Cluster \\
  UT 			& Upstream Tracker \\
  \hline
\end{tabular}
  \caption{
  	Raw Banks used}
  \label{fig:rawbanks}
\end{table}

% ----- 2.1.2 -----
\subsubsection{VP}
\label{ssubsec:vp}

\begin{figure}%
    \centering
    \subfloat[(x,z) layout]{{\includegraphics[width=0.45\linewidth]{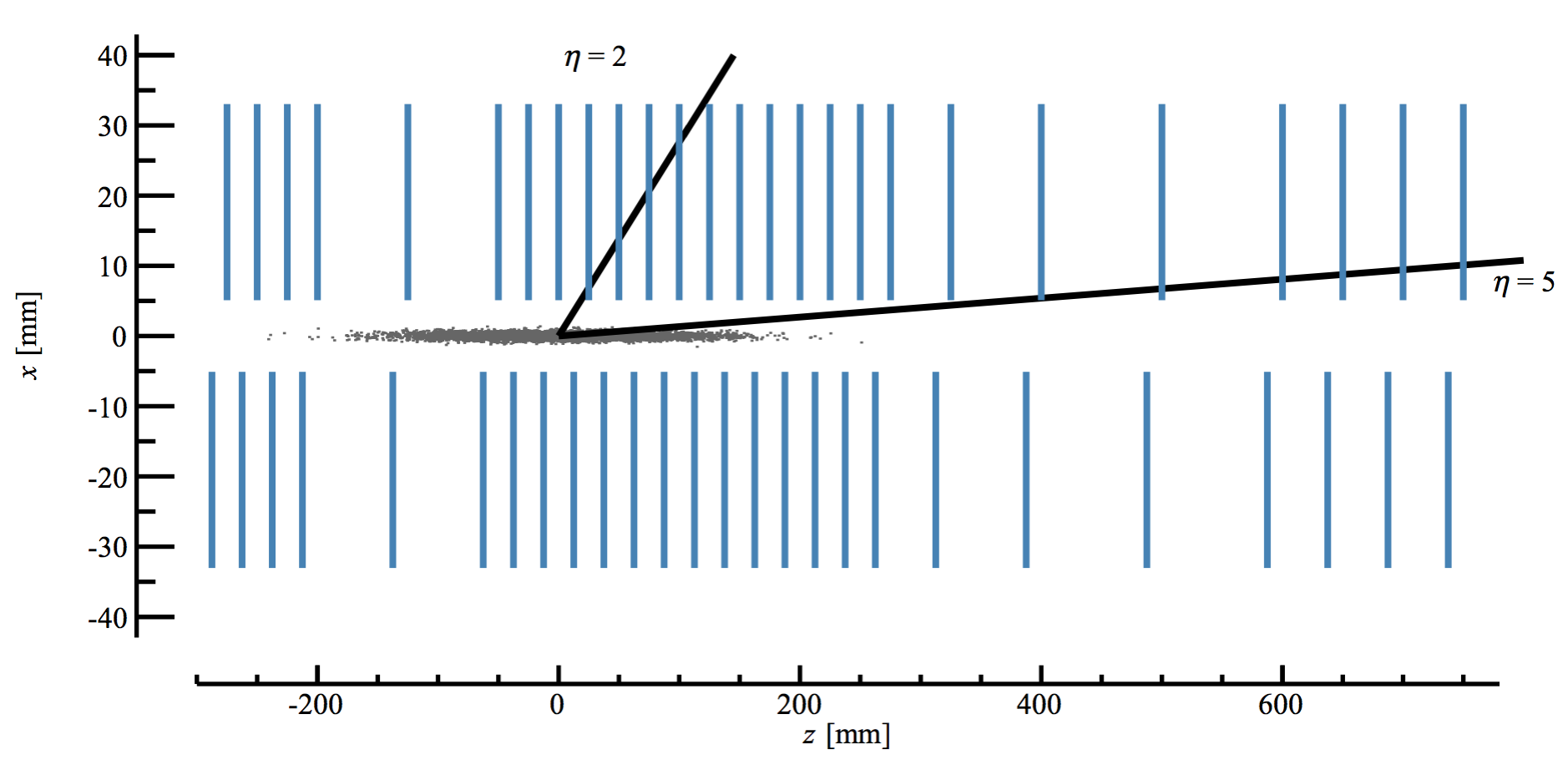}}}%
    \qquad
    \subfloat[(x,y) layout in closed/open positions]{{\includegraphics[width=0.45\linewidth]{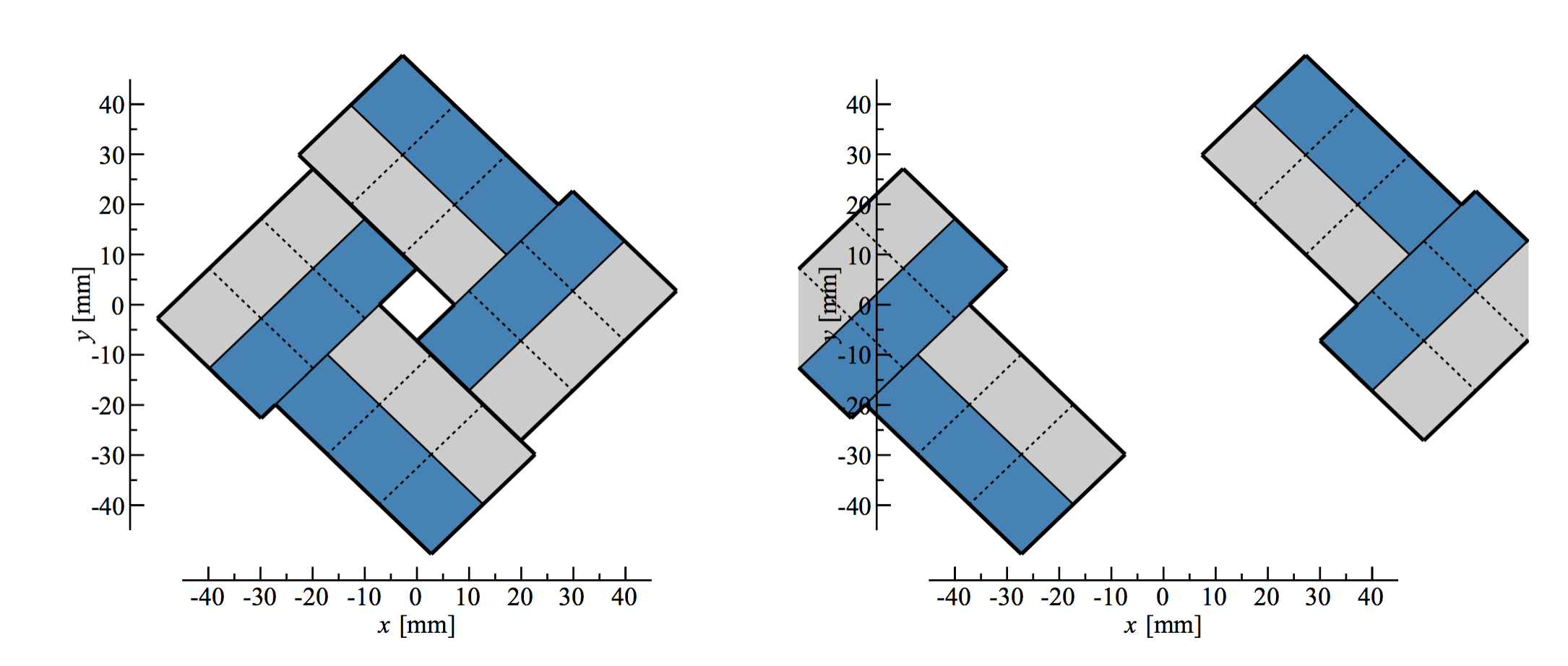} }}%
    \caption{Upgraded VP sensor layout in the \lhcb detector}%
    \label{fig:velo}%
\end{figure}

In the \lhcb upgrade, the Vertex Locator (VELO) will be replaced by a hybrid pixel detector~\cite{LHCb-TDR-013}. The detector, schematically shown in Fig.~\ref{fig:velo} (a), is made of 208 silicon pixel sensors,  arranged as in Fig.~\ref{fig:velo} (b) by groups of 4 and split into 2 sections: A and C. Each group of 4 sensors is called a \textit{module}. As in Section~\ref{ssubsec:rawbank}, we use the size of the raw banks containing the sensor readouts as a proxy for their activity. 

\begin{figure}%
    \centering
    \subfloat[Sample VP readout]{{\includegraphics[width=0.45\linewidth]{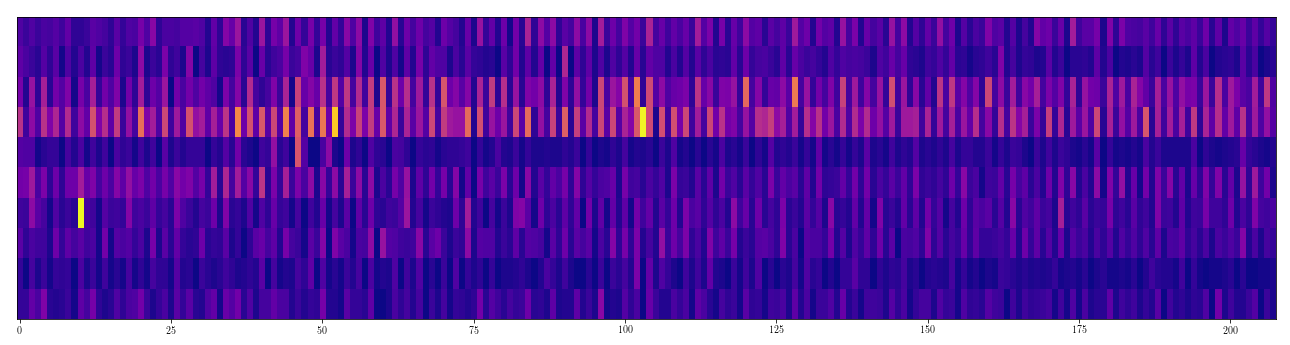} }}%
    \qquad
    \subfloat[Sample max-pooled VP readout]{{\includegraphics[width=0.45\linewidth]{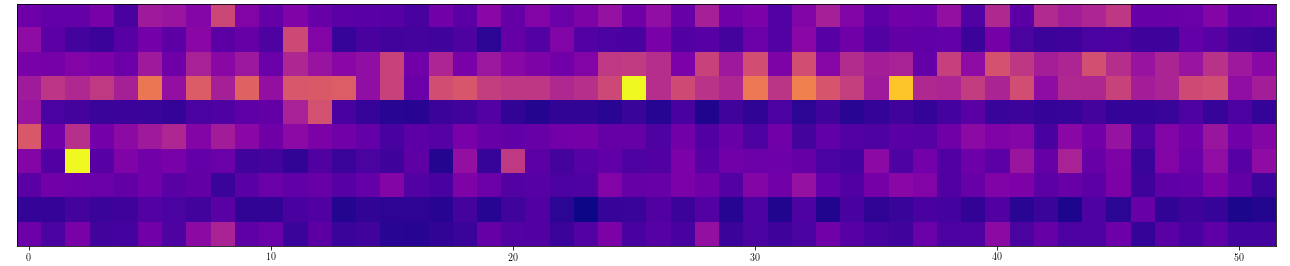} }}%
    \caption{VP readouts: (left) Full readout (right) Max-pooled}%
    \label{fig:vpraw}%
\end{figure}

Note that once the readout values for the 208 sensors are retrieved, we can reduce the dimensionality to work at a module level, either by pooling the maximum sensor values for each module (\textit{max-pool}) or by averaging the sensor readouts across the modules (\textit{avg-pool}). These are common dimensionality reduction techniques, freely inspired from downsampling methods used in Computer Vision~\cite{NIPS2012_4824}. We show the effect of this downsampling on a set of raw events in Fig.~\ref{fig:vpraw} (a) - full readout - and (b) - max-pooled, selecting the maximum readout value for each module.

% ---------------
% ----- 2.2 -----
\subsection{Alternative Global Event Cuts}
\label{subsec:output}

Ultimately, we would like to learn an unbiased (i.e. which does not impact the downstream physics channels) filter which can reduce the rate at which the trigger has to process events downstream, all while preserving as many informative events as possible (i.e. as much signal as possible). This means we would like to find some global values that are as unbiased as possible with respect to the physics but also that are decently correlated to the computational load of the event they represent. If these conditions are met, we can attempt to use these quantities as the basis for Global Event Cuts, provided that they are learnable with the available information described in Section~\ref{subsec:input}.

% ----- 2.2.1 -----
\subsubsection{Number of Primary Vertices}
\label{ssubsec:npv}
The first metric we could try to predict is the number of primary vertices that were either reconstructed (\texttt{Event/pRec/VertexPrimary}) or generated (\texttt{Event/Gen/VertexPrimary}), since we have access to Monte-Carlo ground truth. A discussion comparing the two distributions is included in Appendix~\ref{apx:genrec}. A Primary Vertex (PV) is an estimate of the proton-proton interaction point, which is reconstructed from its tracks in the \velo~\cite{Kucharczyk:1756296}.

This number of primary vertices is shown to be positively correlated to the activity of the detectors (see Fig.~\ref{fig:corr_fes}), so it should be reasonable to produce an unbiased estimate of this quantity from the raw bank sizes. 

\begin{figure}[tb]
  \begin{center}
    \includegraphics[width=0.75\linewidth]{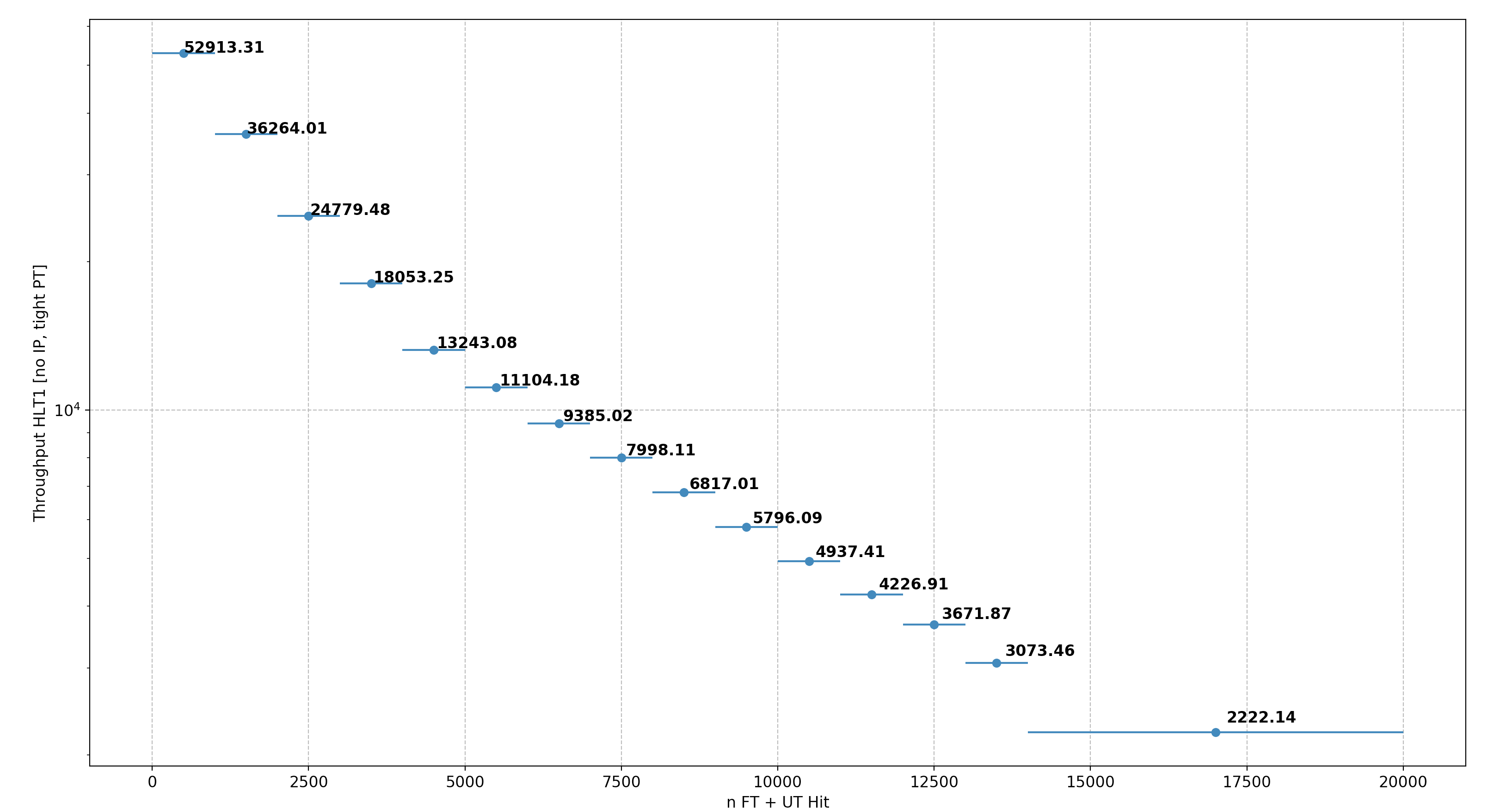}
    \vspace*{-0.5cm}
  \end{center}
  \caption{
  		Throughput as a function of the number of clusters in an event (\emph{Credit: Renato Quagliani})} 
  \label{fig:cost}
\end{figure}

In order for this quantity to be usable as a target, we must also ensure that the number of primary vertices in an event is correlated with the reconstruction cost. Indeed, it seems reasonable to assume that an event with very few PVs will be fast to reconstruct, whereas an event with many PVs will be exponentially more difficult to reconstruct (as it is a combinatorial problem). This trend is reflected in Fig.~\ref{fig:cost}. 

Essentially, this means we can process a lot of low PV-count events but only a few large events. Nevertheless, it is also reasonable to assume that an event with a large number of events will generate a lot of ``information'' (tracks, decays, ...) which means there is a higher probability that it contains interesting physics. Theoretically, each collision is statistically independent, so the more collisions there were, the higher the likelihood of finding a useful decay. This means we will need to find a balance between the large PV-count events (costly but informative) and the low PV-count events (cheap but not as rich). 

% ----- 2.2.2 -----
\subsubsection{\B-decay flagging}
\label{ssubsec:btag}

One of the most important parts of the \lhcb physics program is the study of B mesons. Recognizing the presence of at least a \B -decay in a given event would allow to discard events for which the filter is confident there are none, thereby promoting signal efficiency for this type of decay. Once again, we should be weary of introducing biases here. The learned filter should not treat some decays preferentially: this would be biasing. Some examples to look out for include preferentially recognizing decays with muons compared to decays without, or having a filter accuracy depends on the flight distances of decaying particles.

% ---------------
% ----- 2.3 -----
\subsection{Preliminary analysis}
\label{subsec:prelim}

We begin the analysis by cleaning, analyzing and visualizing the data at hand, forming some initial intuition about what is possible (or not) in the given context. This analysis is made available as a Jupyter Notebook~\footnote{https://gitlab.cern.ch/lhcb-HLT/holistic-information-trigger/tree/master/notebook}.

% ----- 2.3.2 -----
\subsubsection{Feature correlations}
\label{ssubsec:featcorr}

\begin{wrapfigure}{h}{0.35\textwidth}%
\centering
  \includegraphics[width=1.0\linewidth]{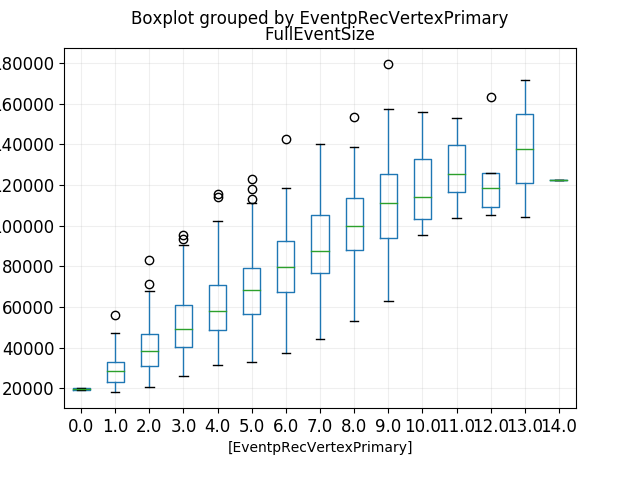}
  \caption{
  		Distribution of the raw event size as function of the number of reconstructed PVs}%
  \label{fig:corr_fes}%
\end{wrapfigure}

\begin{figure}%
    \centering
    \subfloat[Correlation matrix for the \texttt{RawEvent} banks]{{\includegraphics[width=0.45\linewidth]{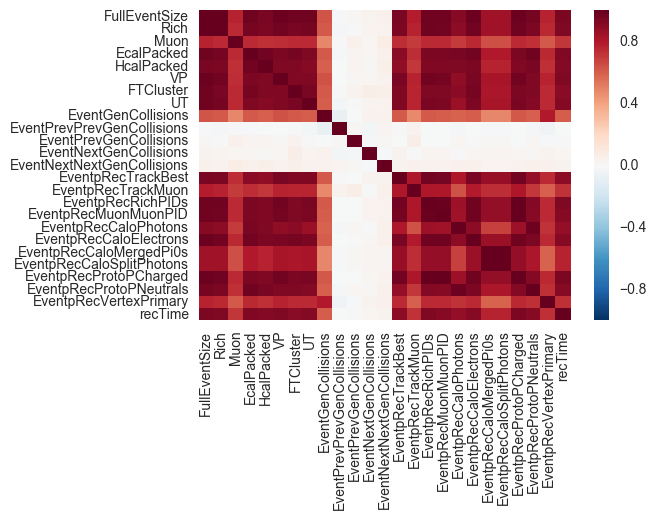} }}%
    \qquad
    \subfloat[Correlation matrix for the raw banks available at training time]{{\includegraphics[width=0.45\linewidth]{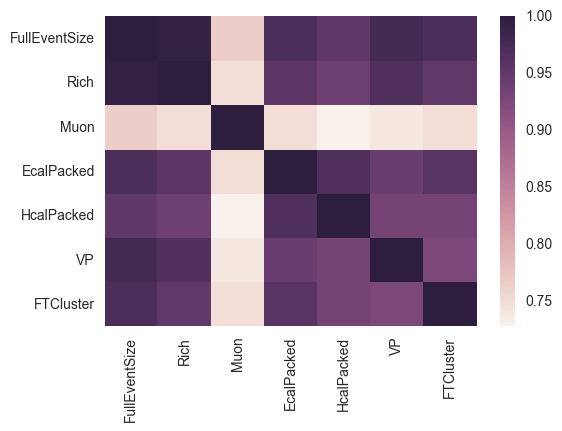} }}%
    \caption{Raw Banks correlation matrix (best viewed in color)}%
    \label{fig:corr}%
\end{figure}

Once the data has been validated and prepared, we can study how it behaves by checking for redundant features or ways to reduce the data, for example by finding high-variance eigenvectors to project onto. In Fig.~\ref{fig:corr}, we show that there is indeed a positive correlation between our input (the raw bank sizes) and our target (the number of reconstructed primary vertices). Most of the inputs are highly correlated, which is to be expected. This also means that the dimensionality could be further reduced. Indeed, a simple Principal Component Analysis (PCA) shows that the first eigenvector holds about $93\%$ of the variance and the second around $5\%$: this means we could reduce from an 8D feature space to a 2D feature space while still retaining $98\%$ of the information content.

Additionally, the fact that some of the inputs are highly correlated can serve as a sanity check, as sensors close together should see similar activity. We observe that the Muon detector has the lowest correlation: it is the last sensor in the forward geometry structure of \lhcb, and we hypothesize that this sub-detector has a higher noise floor than the rest.

We can pursue the correlation analysis by looking at how the number of reconstructed Primary Vertices behaves as a function of a highly correlated feature, which we show in Fig.~\ref{fig:corr_fes} for the \texttt{FullEventSize} raw bank. We know the correlation is high and positive, which is reflected in the positive slope of Fig.~\ref{fig:corr_fes}. Nevertheless we see that the relationship is not straightforward, meaning we will need more than a linear fit to predict the number of PVs given the \texttt{FullEventSize}. The rest of correlation and distribution plots are also available in Jupyter notebooks~\footnote{https://gitlab.cern.ch/lhcb-HLT/holistic-information-trigger/tree/master/notebook}.

% ---------------
% ----- 2.4 -----
\subsection{Learning the mapping}
\label{subsec:learning}

We have established in Section~\ref{subsec:input} a reasonable set of features and in Section~\ref{subsec:output} a set of targets. After having checked the sanity of the data at our disposal in Section~\ref{subsec:prelim} and the presence of correlations in Section~\ref{ssubsec:featcorr} we can proceed to try to learn a mapping to the target, to be potentially used as a global event cut variable.

% ----- 2.4.1 -----
\subsubsection{Multi-Variate Analysis}
\label{ssubsec:mva}
A common set of tools for data studies is Multi-Variate Analysis (MVA), which aims to reduce a large number of variables to a subset of explainable factors. These are routinely used within \lhcb analysis, notably through the TMVA framework~\footnote{http://www.ep.ph.bham.ac.uk/twiki/bin/view/General/TMVA}. We will start by using these standard tools, as they are well understood and represent a sane baseline to eventually compare other ideas to.

These tools include two main classes of algorithms: \textit{random forests}, which learn a set of uncorrelated decision trees, minimizing error variance but limiting the effect on error bias, and \textit{boosting algorithms}, which iteratively compose weak learners, reducing bias but subject to higher error variance~\cite{louppe2014understanding}. These methods provide the additional advantage of being interpretable in the sense that they provide a measure for the importance of each input feature in its prediction. This can be an interesting indicator, serving as a sanity check or as a way to understand the prediction.

% ----- 2.4.3 -----
\subsubsection{Machine Learning and Neural Networks}
\label{ssubsec:NN}
We have touched on the use of data-driven methods, their potential and drawbacks in previous sections. In particular, deep neural networks have been used as powerful building blocks for learning complex non-linear mappings. Their fully differentiable nature allows for end-to-end optimization: powerful features are automatically learned and composed to learn an optimal predictor. They also present a major advantage in a setting such as the trigger, where computations are expensive: while training a neural network is costly, the inference complexity is constant and it can be easily parallelised (since weights are frozen at inference, predictions are a simple forward pass: a sequence of matrix multiplications passed through non-linearities). This could allow the delegation of the heavy-lifting to a large offline system (as is done since Run 2, with the HLT2 running offline on the farm), and later allow the algorithm to run online. This is of course appealing in the context of the upgrade which has the ambition of implementing a fully software-based trigger, doing away with the hardware L0 trigger.

% ----- 2.4.4 -----
\subsubsection{Attention Mechanisms}
\label{ssubsec:attention}
Recurrent models have seen great success in dealing with sequence-like data~\cite{VaswaniSPUJGKP17}, most notably in fields like translation~\cite{BahdanauCB14}. Traditional methods will pass the word to be translated and $n$ of its neighbours, which serve as context, into the network which in turn outputs a prediction. This fixed context is not ideal as sentences are an example of weakly ordered sequences: while order does matter it is not a fixed relationship. Dependencies can be long range, with words at the beginning of a sentence giving context to a word further downstream. This means the direct neighbourhood of a word does not necessarily reflect its actual context, and might miss these long range dependencies.\\

The idea of an attention mechanism has been proposed to remedy this issue. The basic idea is to \textit{learn} a context vector - a weighted sum of the entire input sequence. The weight of each input element would represent its importance in the prediction made by the network downstream. Since the assignment of attention is done with a \textit{soft-max} assignment (compared to a non-differentiable \textit{arg-max}~\cite{2018arXiv180204223N}), gradients are allowed to propagate through the network. This means attention is learned end-to-end, at the same time as the features that optimize the desired loss function: the network learns where to ``look'' and how to compose the entries it attends to for an optimal prediction. This mechanism allows a straightforward modelling of long-term dependencies, but also provides a method to handle loosely ordered sequences, a long-standing problem in the sequence modelling field~\cite{2015arXiv151106391V}. Note that this idea is related to the idea of attention in human vision, where a central point is in focus, helped by an outside ring which is blurry but still informative, as well as the composition of several focus points over time helping to analyze the scene~\cite{abs-1109-3737}.

An example of loosely ordered sequences actually appears in our dataset, in the set of 208 VP sensors. The sensors are unevenly distributed spatially (see Fig.~\ref{fig:velo} (a)), and have different orientations (split into 2 sections of 104 sensors, and in groups of 4 modules, see Fig.~\ref{fig:velo} (b)). Nevertheless, for a given track (which, modulo some scattering, is linear in this part of the \lhcb detector), there is a coherent sequence of sensors being fired: this sequence is not ordered in the sense of the sensor index, but the particle necessarily follows a linear path and hence forms a coherent sequence of hits. This motivates the use of attention mechanisms, which we designed to model these long-range and non-index obvious sequences.\\

The attention mechanism is mostly used in sequence-to-sequence models, as in the successful translation example~\cite{BahdanauCB14}. Our setting however is purely feed-forward: we want to classify or regress upon a value from an input sequence - in this case the VP readouts. Luckily, the context vector (the weighted sum of attended inputs) is of constant size for a given sequence. This means it can be passed as input to any machine learning algorithm as input, in the form of a vector of arbitrary dimension encoding a compressed representation of the sequence and its context. This idea was developed in recent work, which showed that an attention mechanism is able to remember the importance of elements several thousand elements away in a sequence for a long-range arithmetic task~\cite{RaffelE15}. This was a difficult task for classic LSTM-based methods of encoding sequences~\cite{Hochreiter} as the gradients would fade after observing too many elements of the sequence, and the hidden state would then ``forget'' which value was important for its prediction.

As an added motivation for the use of a feed-forward formulation, it has been shown that they offer more stability in training compared to recurrent models, are easily parallelisable~\cite{GehringAGYD17} (given that there are no sequential dependencies to the output) and usually offer a speedup at inference time~\cite{abs-1211-5063}. While it is commonly believed that recurrent models are more expressive, recent work has also shown that this assumption often does not hold in the real-world~\cite{KakadeLSV16}, with some feed-forward architectures outperforming recurrent formulations on a variety of baselines~\cite{abs-1803-01271}, and even work hypothesizing an equivalence in practice between recurrent models and feed-forward architectures~\cite{abs-1805-10369}, noting for example that most recurrent models are trained with a fixed-size history.\\ \\

We propose to adapt this feed-forward attention scheme for our purposes. To do so we present a reusable PyTorch~\cite{paszke2017automatic} library which implements the ideas introduced in~\cite{RaffelE15}. This library is released as an open source project on Github \footnote{https://github.com/dtsbourg/ff-attention}. In it we successfully reproduce the experiments proposed in the paper, and extend the functionality to our own needs. This usage is detailed in Section~\ref{subsec:ffa}.\\ \\ \\

\begin{figure}[hb]
  \begin{center}
    \includegraphics[width=0.75\linewidth]{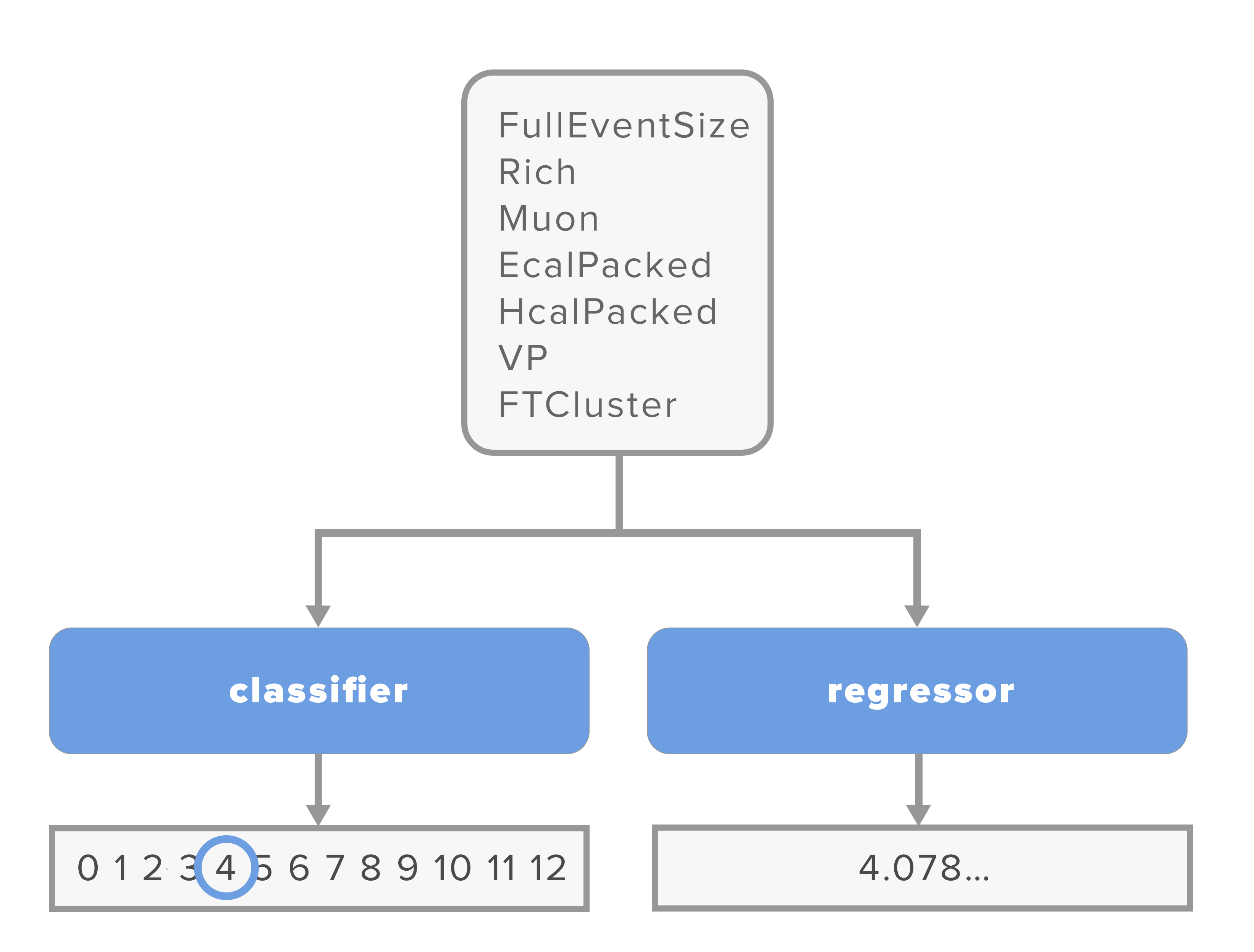}
    \vspace*{-0.5cm}
  \end{center}
  \caption{
  		Schematic pipeline for classification (left) and regression (right) of the number of Reconstructed PVs from Raw Bank Sizes}
  \label{fig:classifier}
\end{figure}

\newpage
\section{Experiments}
\label{sec:experiments}

In the following experiments we use an Upgrade minbias sample~\footnote{Dirac path: \texttt{/MC/Upgrade/Beam7000GeV-Upgrade-MagDown-Nu7.6-25ns-Pythia8/\newline Sim09c-Up02/Reco-Up01/30000000/LDST}}, containing $26\,770$ minbias events. The raw bank sizes are obtained by running a custom \davinci script. Details for reproducing the analysis are available in the \lhcb Gitlab repository~\footnote{https://gitlab.cern.ch/lhcb-HLT/holistic-information-trigger.git}. In all training instances, the data is split into a training dataset and a testing dataset, with an $80/20$ split. All experiments are cross-validated.

\subsection{Classifying the number of Reconstructed Primary Vertices from Raw Bank sizes}
\label{subsec:classification}

\begin{wrapfigure}{R}{0.5\textwidth}
\centering
    \includegraphics[width=1.0\linewidth]{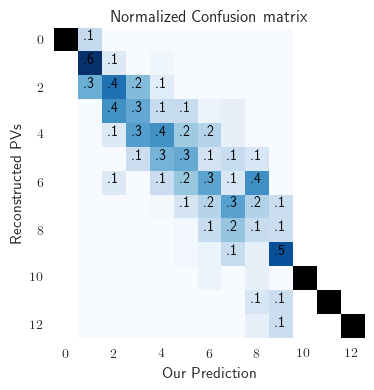}
  \caption{
          Confusion matrix in predicting the number of Primary Vertices from Raw Bank Sizes}
  \label{fig:confusion}
\end{wrapfigure}

The number of reconstructed primary vertices is inherently a discrete value, so it makes sense to consider the problem of predicting the number of reconstructed primary vertices as a multi-class classification problem, wherein we want to place an event into one of the ``bins''. One fundamental issue with this approach is that there is no way to know a priori the number of bins. We must have a large enough training sample to ensure that all bins contain examples and that the prediction is not limited to an erroneous range.

The idea here is to use the size of the raw banks (described in Table~\ref{fig:rawbanks}) as input features for a classification algorithm. The proof of concept is run with a simple MVA, as described in Section~\ref{ssubsec:mva}. We present a Random Forest Classifier from the \texttt{scikit-learn}~\cite{scikitlearn} standard toolbox, but several other MVA methods were tested in development - ignored here for brevity. The pipeline is presented schematically in Fig.~\ref{fig:classifier}.

The Random Forest classifier we obtain after training is, as can be expected, biased towards the mean - underestimating busy events, overestimating quiet events. Nevertheless the error is Gaussian (hence predictable), with an error $\sim \mathcal{N}(\mu=-0.172,\,\sigma^{2}=1.1412)$. We can also plot a confusion matrix, which presents a condensed version of predictions versus ground truth. It is shown in Fig.~\ref{fig:confusion}, and confirms the idea that the predictor is unbiased on average, but does present variations at the extrema of the target distribution (very large or small number of PVs).

\begin{figure}[hb]
  \begin{center}
    \includegraphics[width=0.95\linewidth]{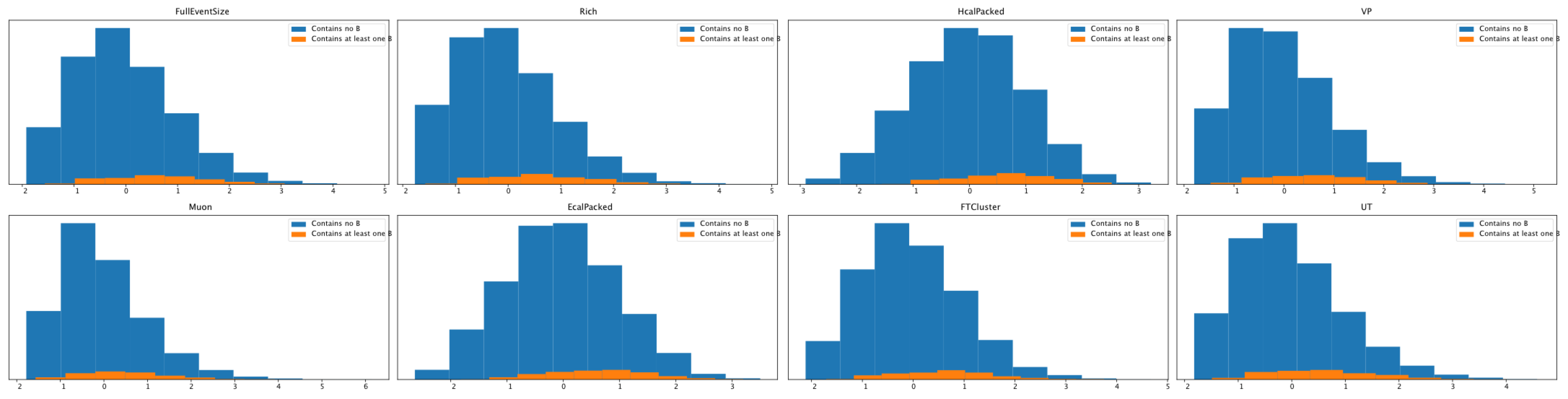}
    \vspace*{-0.5cm}
  \end{center}
  \caption{
          Feature distributions for \B and non-\B containing events}
  \label{fig:bfeat}
\end{figure}

\newpage
\subsection{Flagging \B-decays}
\label{subsec:btag}

As in other analyses, we start by looking manually for correlations or obvious patterns. We show the distribution of the raw bank sizes in Fig.~\ref{fig:bfeat}. It reveals the fact that the classes are extremely imbalanced (as mentioned in Section~\ref{ssubsec:btag}), but also that they seem to overlap in the feature space. We include a projection of the manifold in the Appendix~\ref{apx:manifold}, Fig.~\ref{fig:manifold}. This suggests that learning a clear hyperplane separating the events containing at least a \B-decay from those that don't will prove challenging for any binary classifier. Fortunately, we do have the relaxed constraint that we want our cut to be efficient, not perfect, softening the classification loss.

\subsubsection{Polytope cut}
\label{ssubsec:polycut}
As shown in Fig.~\ref{fig:bfeat}, the features for \B and non-\B containing events overlap heavily. We could, however, envision a simple cut that takes the soft extrema of events which contain \B-decays and removes all events outside these boundaries. As we apply this for each feature we essentially hope that the data can be separated by hyperplanes in the feature space. Unfortunately, this is not the case here, as we do not obtain any significant gains without substantial efficiency losses, not to mention the lack of statistical robustness of this simple polytope cut.

\subsubsection{Binary classification}
\label{ssubsec:binaryclass}
Binary classification is a matter of separating two distinct classes based on their features. The assumption is that the features present sufficient statistical divergence that they are separable. We test this assumption in our context, where we apply a 4-layer feed-forward neural network with a standard configuration for binary classification (Binary Cross-Entropy loss).\\

This does not show any positive results though: an Area Under the Curve (AUROC) of $0.51$, yielding a throughput gain of only $0.61\%$ (for no \B-decay loss). The hypothesis that the manifolds are heavily connected even in the feature manifold is again comforted. This result is not entirely surprising: the very small amount of \B-decays even within an event that does contain one (decays yield a large number of other particles apart from \B mesons) coupled with the fact that there are only a fraction of these events that actually contain \B-decays ($\sim \frac{1}{20}$) makes the learning problem more like searching for a needle in a haystack than a clear-cut binary classification between two distinct classes, essentially meaning a \textit{lot} more data is needed to learn the mapping.

\subsubsection{Oversampling and weighted losses}
\label{ssubsec:weighted}
In an attempt to fight the inherent class imbalance (around $6.2\%$ of events contain \B-decays in our sample, which amounts to approximately $1\,700$ events out of $26\,770$), some best-practice methods were implemented. Notably over- and under-sampling the imbalanced classes usually helps neural networks treat positive and negative classes equally. Note that this biases the estimator - the expectation with respect to the negative examples is not equal to the expectation on the full dataset. This is a common issue with negative sampling~\cite{MikolovSCCD13}.  

The dual formulation of over-/under-sampling is to weight the loss for either the positive (presence) or the negative (absence) class to balance the gradient updates from both classes: this helps fix the imbalance without skewing the input distribution.

In practice, neither methods seemed to yield satisfactory results, amplifying the fact that events with and without \B-decays are difficult to distinguish, at least with the data at our disposal.

\subsubsection{Ranking}
\label{ssubsec:rank}
In light of the lack of success of binary classification methods, given that the classes are not easily separable and that the data at hand is not discriminatory enough, we attempted another relaxation of the problem, making the classification problem even softer. The idea is to train a network to estimate the likelihood of an event to be part of a class or not, with the constraint that the likelihood should simply be higher for the correct class. By applying this ranking formulation, we can set the problem as a pairwise decision process: given two events, which one is most likely to be the event containing a \B-decay?

Again due to the lack of data, no conclusive results could be obtained from this formulation.\\ \\

\begin{figure}[h]
  \begin{center}
    \includegraphics[width=0.75\linewidth]{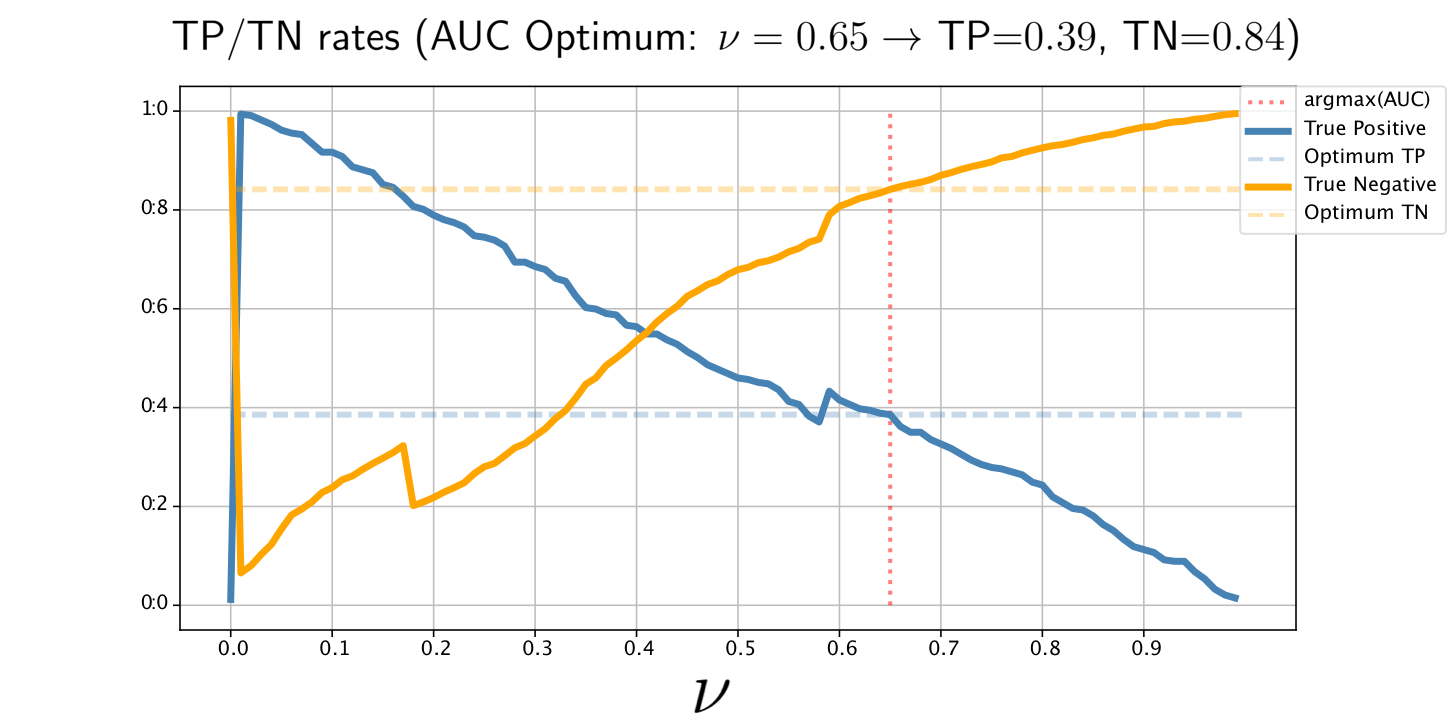}
    \vspace*{-0.5cm}
  \end{center}
  \caption{
          True Positive/Negative Rates at max-AUC (=0.6) with Raw Bank features}
  \label{fig:oneclass}
\end{figure}

\subsubsection{Anomaly detection / One-class learning}
\label{ssubsec:anomaly}

As we've shown in the previous experiments, it is difficult to cleanly separate the two classes of events. Fortunately, the main objective here is to retain the maximum number of events that actually contain a \B-decay, while discarding as many events with no \B-decay as possible: we want to maximize throughput while minimizing signal loss. In other words, recall is more important than precision here. Recall is not a differentiable objective, and some approaches have attempted to solve this issue by optimizing precision at a fixed recall rate~\cite{2016arXiv160804802E}. We decide to use a more straightforward way to go about this is to apply methods inspired by anomaly detection.

\begin{figure}[h]
  \begin{center}
    \includegraphics[width=0.8\linewidth]{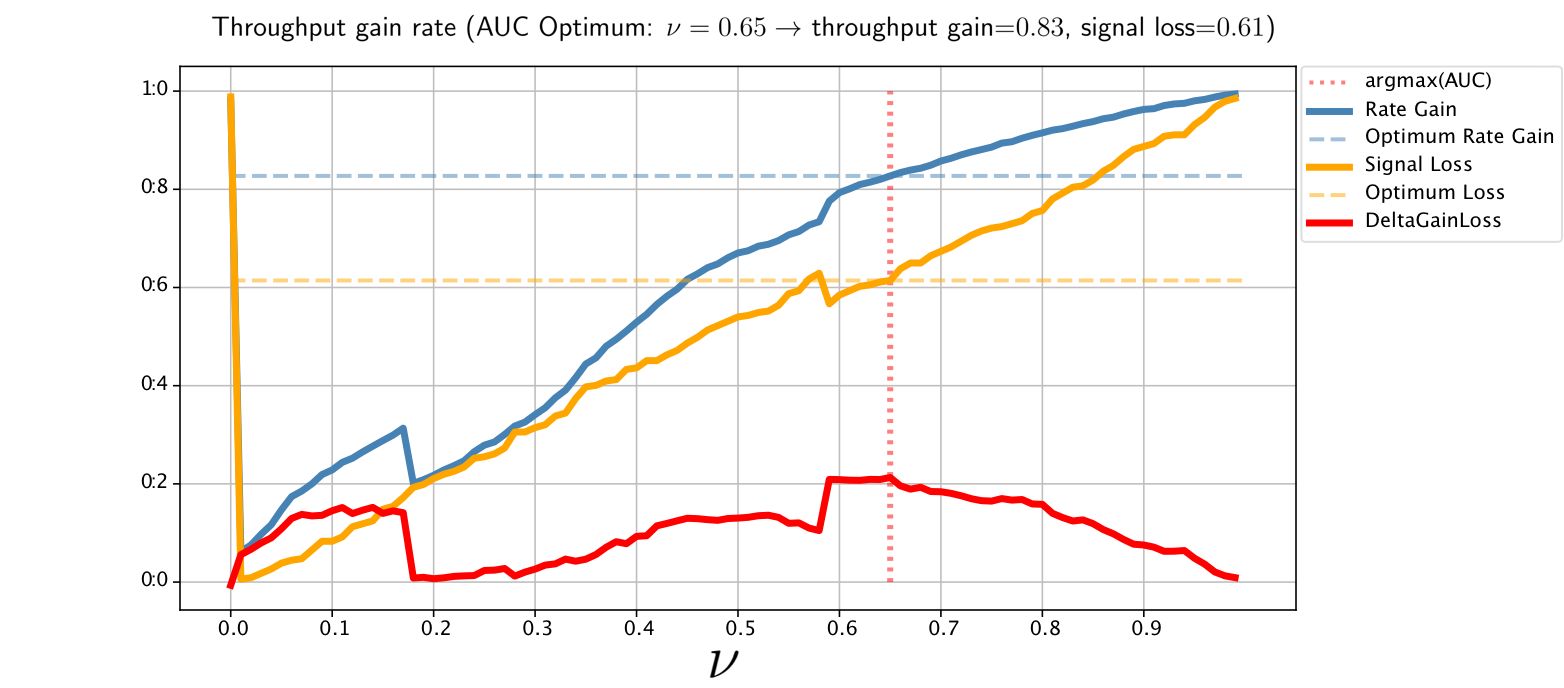}
    \vspace*{-0.5cm}
  \end{center}
  \caption{
          Throughput Gain and Signal loss with One-Class SVM (Raw Bank features)}
  \label{fig:sigloss}
\end{figure}

In binary classification, the assumption is that the two classes are separable, whether in the feature space or lifted with a kernel. Anomaly detection instead formulates the problem as a one-vs-all situation: one class models nominal behavior, and anything that deviates from it would be considered an anomaly, an outlier. In many cases where this setting is applied, the nominal behavior class is still separable, but separable from any other class but itself. This is useful because we only have to model what a positive example looks like. In our case, this would be modeling events containing a \B. This difference is illustrated in Fig.~\ref{fig:anom} (Appendix~\ref{apx:nusvm}).

\begin{figure}[h]
  \begin{center}
    \includegraphics[width=0.75\linewidth]{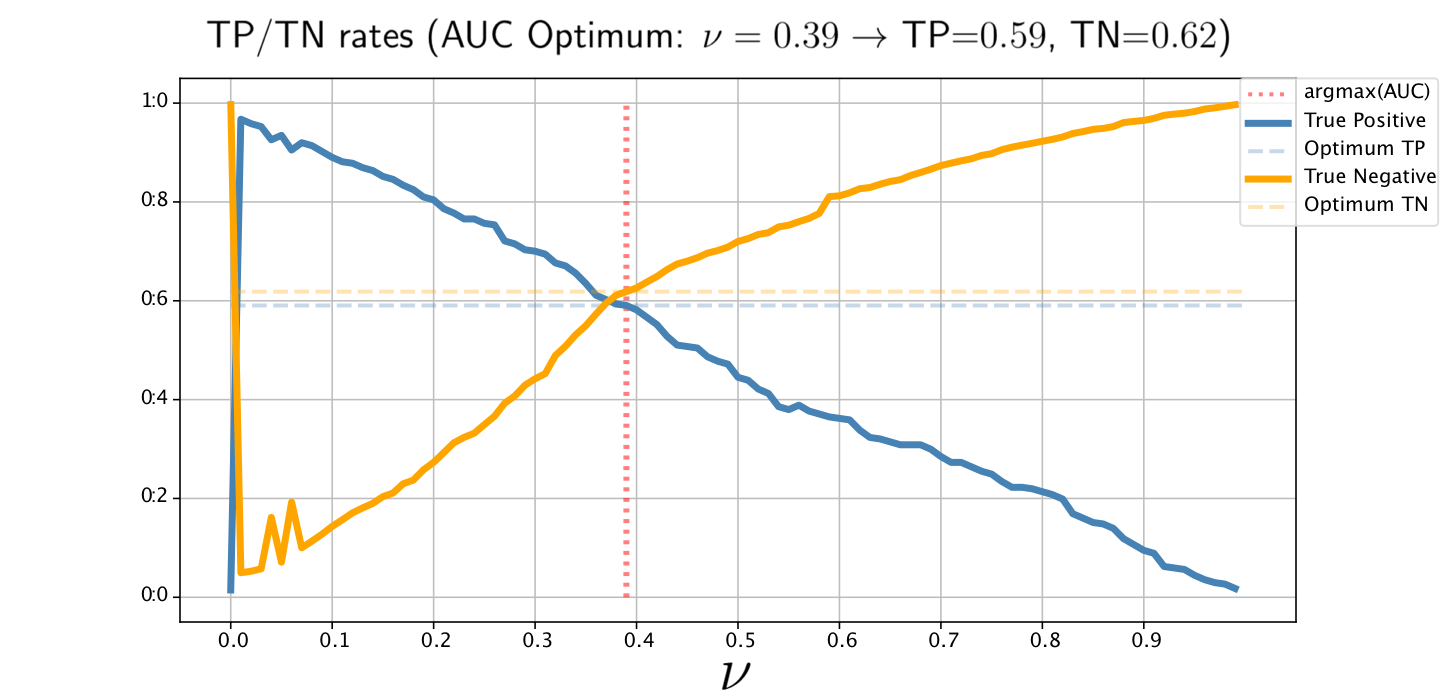}
    \vspace*{-0.5cm}
  \end{center}
  \caption{
          True Positive/Negative Rates at max-AUC (=0.6) with VP features}
  \label{fig:oneclass_vp}
\end{figure}

In our setting though, we have seen that separability is difficult to achieve. To solve this issue, we look to a parametric model which can control how strongly outliers are rejected. In particular, One-Class Support Vector Machines (SVM)~\cite{Scholkopf} learn a soft boundary around the training data, where the ``softness'' can be controlled by a hyper-parameter $C$. Formally, it is controlled by a parameter $\nu = \frac{1}{2C}$, which represents the fraction of training errors (and a lower bound on the number of support vectors). 

This parameter can be tuned to determine how tightly we model the nominal (i.e. \textit{presence of \B}) behavior: if the boundary is forced to include every single training event with \B ($\nu \rightarrow 0$), then only a small fraction of non-\B events will be placed outside this boundary. We preserve most of the signal events but don't get much gain in terms of throughput. On the other hand, if we are willing to sacrifice some signal, we can also discard more background ($\nu \rightarrow 1$). In essence, $\nu$ controls the balance between precision and recall, essentially balancing the throughput gain and the loss of signal. For a more intuitive view of $\nu$'s influence, we refer the reader to Appendix~\ref{apx:nusvm}, Fig.~\ref{fig:nu_infl}.

\begin{figure}[hb]
  \begin{center}
    \includegraphics[width=0.8\linewidth]{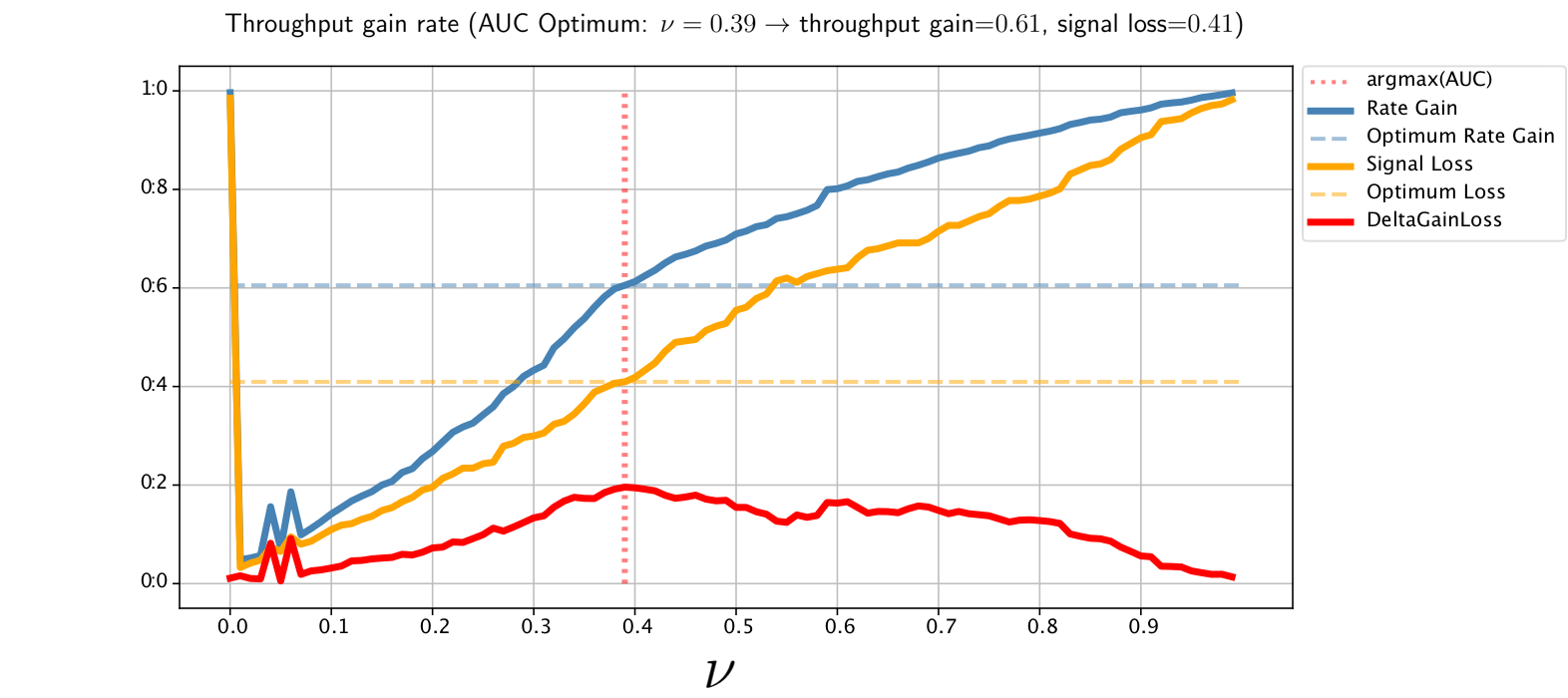}
    \vspace*{-0.5cm}
  \end{center}
  \caption{
          Throughput Gain and Signal loss with One-Class SVM (VP features)}
  \label{fig:sigloss_vp}
\end{figure}

In Fig.~\ref{fig:oneclass} and~\ref{fig:sigloss}, we present the results when training a One-Class SVM with a sigmoid kernel on the raw bank features. The discussed precision/recall trade-off is equivalent to finding the right balance between True Positive (TP) and True Negative (TN) rates (shown in Fig.~\ref{fig:oneclass}). In our case, the TP rate is the proportion of events containing a \B-decay that are correctly set inside the boundary, and the TN rate is the proportion of events not containing any \B-decay that are placed outside the boundary. \\

If we consider the classifier decision boundary as our cut, the throughput gain is given by the total number of examples classified as negative, i.e. the sum of events incorrectly classified as containing \B-decays and the events actually containing a \B-decay being discarded. Given the class imbalance, the True Negative rate serves as a close enough approximation of the rate gain. \\

In a perfectly separable problem, which as we have seen is not the case here, we would have TP$=1$ and TN$=1$. Also, since the ``usefulness'' of a given event is ill-defined, there is no global optimum. We must strike a balance in terms of the rate at which events that actually contained a \B are discarded and the rate at which events with no \B-decays are discarded. In other words, we must balance the throughput gain and the signal loss, which is defined as the number of events containing \B-decays that are being discarded (this is the False Negative Rate).

One possibility is to choose the maximum Area Under the Curve (AUC), in this case shorthand for the AUROC (for Area Under Receiver Operating Characteristic curve), which measures the probability that our classifier will rank a random positive example higher than a random negative example~\cite{aucroc}. This is an equivalent formulation to our Ranking Problem, discussed in Section~\ref{ssubsec:rank}. This measure makes sense because it serves as a summary statistic, encapsulating sensitivity (or True Positive Rate / Recall) and False Positive Rate (which captures how many negative examples are considered as positive, i.e. in our case how many non-\B events are classified as \B by the classifier). We show that the maximum AUC point is also the point with the maximum gap between throughput gain and signal loss (see Fig.~\ref{fig:sigloss} and ~\ref{fig:sigloss_vp}). At this optimum ($\nu=0.65$) we report a throughput gain of $0.84$ (which means we reduce the total rate by around $84\%$), with $39\%$ signal efficiency.\\

Note that other local optima can be found. If we decide to value the preservation of \B samples much higher for example, we could choose a point at $\nu = 0.17$ which allows us to keep $\approx 90\%$ of \B events while discarding about $1$ in $4$ non-\B events. This results in a total throughput gain of around $28 \%$, which is much lower than the previous optimum, but with a much lower signal loss, only around $12 \%$.\\

We also run the analysis on the VP sensor readouts, a data-source detailed in Section~\ref{ssubsec:vp}. With similar settings, we report a similar maximum AUC value of $0.602$ at $\nu=0.39$, shown in Fig.~\ref{fig:oneclass_vp} and Fig.~\ref{fig:sigloss_vp}. Here we obtain a max-AUC optimum which improves the throughput by $62\%$ with $41\%$ signal efficiency.\\

Another advantage of this parametric formulation is that the parameter $\nu$ will also control the inference-time complexity of the classifier. Indeed, the test-time complexity of an SVM classifier is $\mathcal{O}(n_{SV} \cdot d)$ with $n_{SV}$ the number of Support Vectors and $d$ the dimensionality of the problem~\cite{6796736}. Since $\nu$ controls the number of support vectors, it can be used as a proxy for the inference complexity as well.

\subsection{Predicting the number of Reconstructed Primary Vertices from \velo Pixel activity}
\label{subsec:regression}
In this experiment, we focus solely on one of the detectors' readouts. Instead of taking the total sizes of the raw banks for each sub-detector (Table~\ref{fig:rawbanks}), we focus only on the readouts from the \velo Pixel detector, whose architecture and readouts are described in Section~\ref{ssubsec:vp}. 

\begin{figure}[tb]
  \begin{center}
    \includegraphics[width=0.5\linewidth]{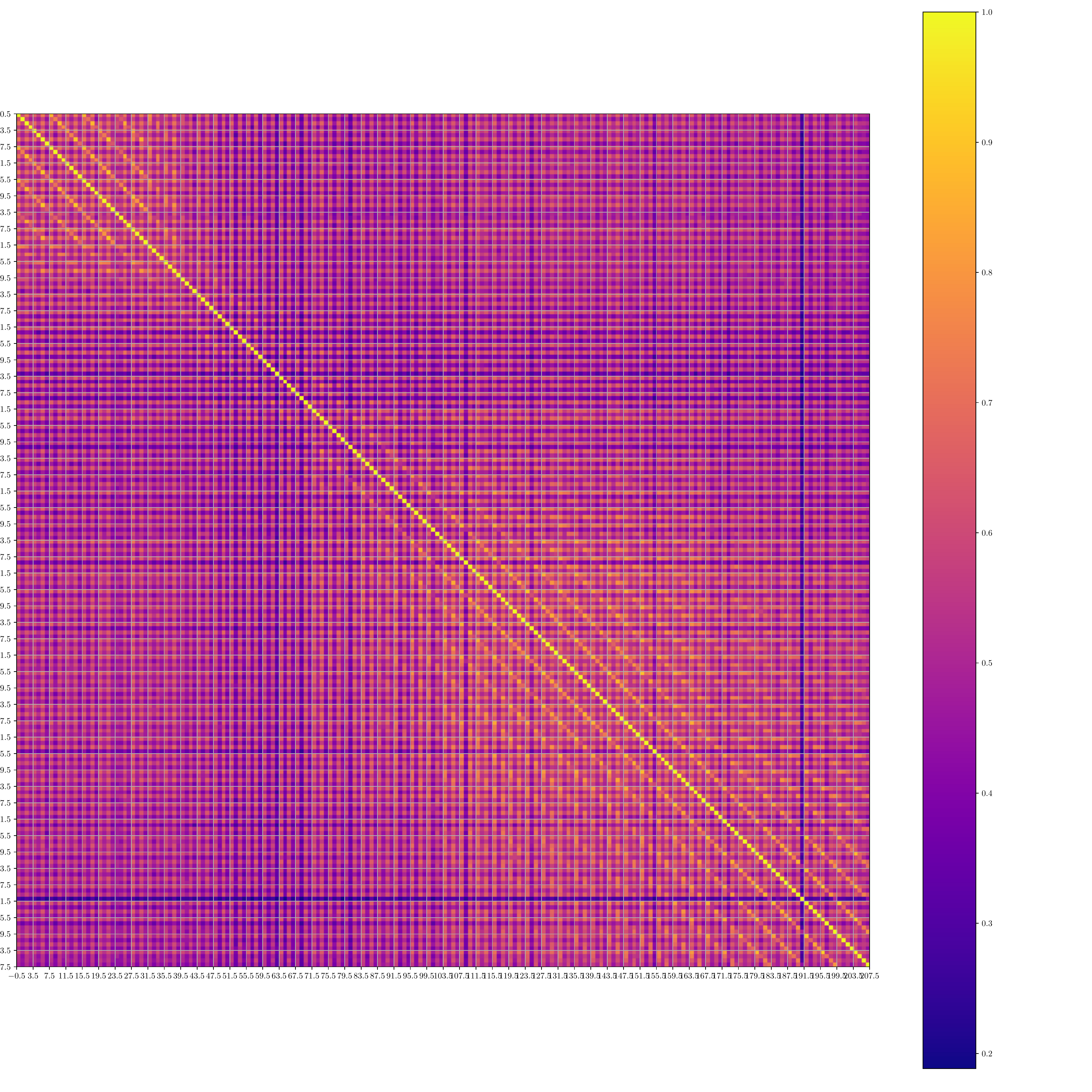}
    \vspace*{-0.5cm}
  \end{center}
  \caption{
          Correlation Matrix of the VP sensors (best viewed on a monitor)}
  \label{fig:correlation}
\end{figure}

Some interesting patterns appear in Fig.~\ref{fig:correlation}, which represents the correlation between each of the 208 \velo Pixels. We can guess the beamspot thanks to the two more correlated areas (upper left and bottom right on Fig.~\ref{fig:correlation}) which match the geometry of the detector (see Fig.~\ref{fig:velo}). We also make out the submodule structure, with the 4 sensor groups appearing correlated by pairs (one pair on either side). This even uncovers some interesting phenomena like large outliers picked up by the sensor (e.g. photons radiating a single module like \texttt{\#117} in Fig.~\ref{fig:correlation}). The analysis of this outlier is provided in the Gitlab repository as well.\\

The main advantage of a data-driven approach is that we can directly apply the same algorithms, given a new fit, to different data. We want to proceed in a data-agnostic way: the hypothesis is that the less domain knowledge is baked into the algorithm, the less chance for bias in the results. This, of course, is an arbitrary and debatable argument but holds its ground in this context.

Unfortunately, the results do not get much better when applying these methods. We achieve similar accuracy with the \velo Pixel sensors than with the rest of the detector information. This behavior is expected as we have seen that the VP sensor activity is, on average, highly correlated to the rest of the sub-detectors' activity. In consequence, we will try a new method in Section~\ref{subsec:ffa}, leveraging the pseudo-sequential structure of the \velo Pixel readouts.

\subsection{Feed-Forward Attention Mechanisms}
\label{subsec:ffa}
Let $x_t$ be the input sequence defined by the activities from each of the pixel sensors, ordered with the index numbers given by the VP detector architecture (as in Fig.~\ref{fig:velo}). They can also be pooled into a $52 \times 4$ matrix representing the 4 pixels per module. Notation will remain unchanged as the logic is the same, only the sizes will have to be updated at implementation time. Both configurations are implemented and tested in the repository.

We start by encoding the input sequence with a feed-forward neural network $h_t = f(x_t)$. This is simply a transformation of the input sequence, taken as a vector, to a $H$-dimensional embedding ($H=104$ in our experiment). In this case $f$ is a 3-layer ($208 \times 208 \times 208$) feed-forward neural network with Leaky ReLU activations~\cite{XuWCL15}, but can be replaced by any embedding function - even the identity.

The context vector, which is the desired compressed representation of the sequence, is a weighted sum of the embedding $h_t$, where the weights are the individual importances of each element of the sequence. These are computed through the softmax function which assigns probabilities to each class based on the output of a function. In this case we get the attention probabilities $\alpha_t = softmax(e_t)$ where $e_t = \sigma(W \cdot h_t + b)$ is the output of the embedding through a non-linear activation layer - we use $tanh$ - and $softmax(x)_i = \frac{\exp{x_i}}{\sum^H \exp{x_j}}$. The context vector is then given by $c = \sum \alpha_t \cdot h_t$.

This context vector $c$ is then passed to the final output layer. In this experiment, we use another simple feed-forward network (3-layer, $208 \times 104 \times 52$) whose purpose here is predicting the desired quantity, for example the number of reconstructed PVs in the given event. Note that here also the output function can be replaced at will: several experiments have shown the efficiency of passing automatically learned features (through ANNs) into traditional Machine Learning algorithms~\cite{DonahueJVHZTD13}.\\ 

All the networks are implemented in PyTorch. They are initialized with the Xavier~\cite{Glorot10understandingthe} method, and optimized with the Adam~\cite{KingmaB14} optimizer (learning rate $=0.0001$). We apply batch normalization~\cite{IoffeS15} in each hidden layer, as well as dropout~\cite{srivastava14a} (rate $= 0.3$) to avoid over-fitting.\\ \\

\begin{figure}[tb]
  \begin{center}
    \includegraphics[width=1.0\linewidth]{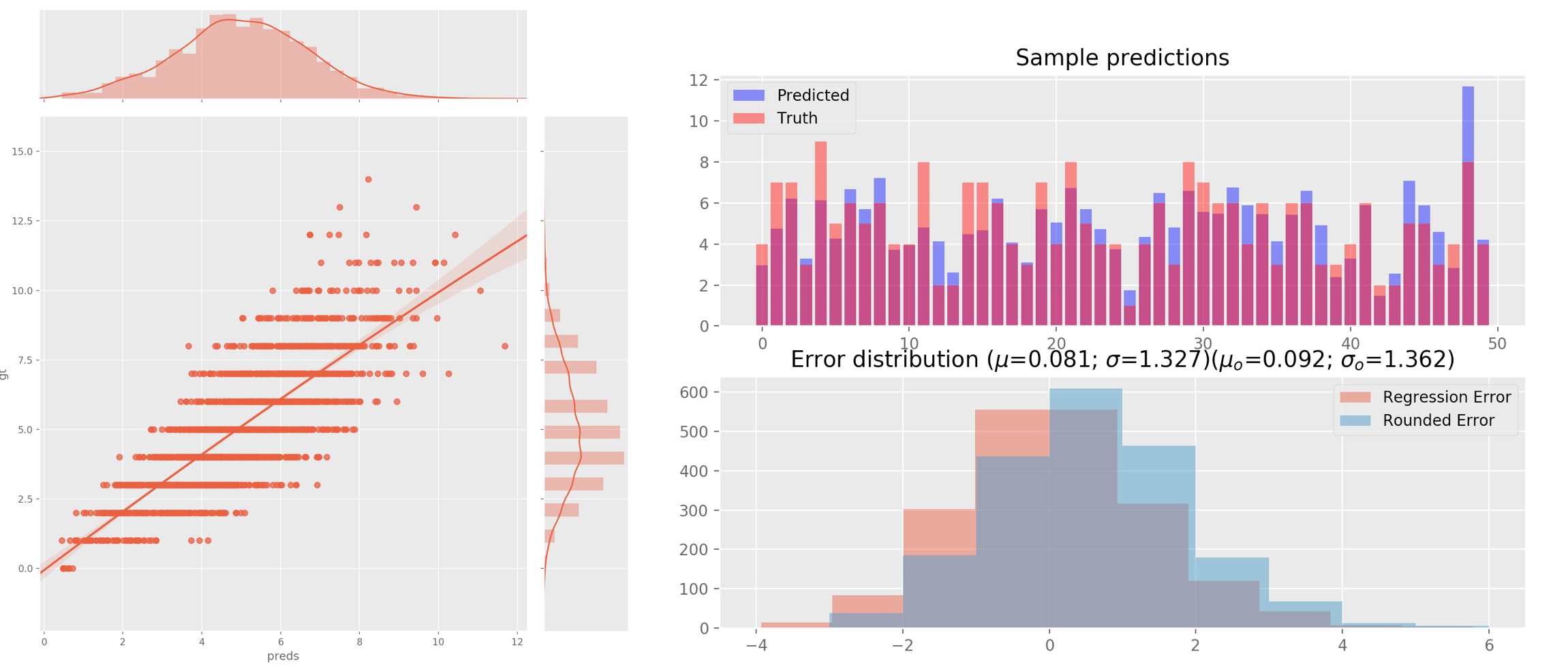}
    \vspace*{-0.5cm}
  \end{center}
  \caption{
          Predictions and error distributions on the number of reconstructed Primary Vertices}
  \label{fig:ffa_preds}
\end{figure}

\begin{figure}[tb]
  \begin{center}
    \includegraphics[width=1.0\linewidth]{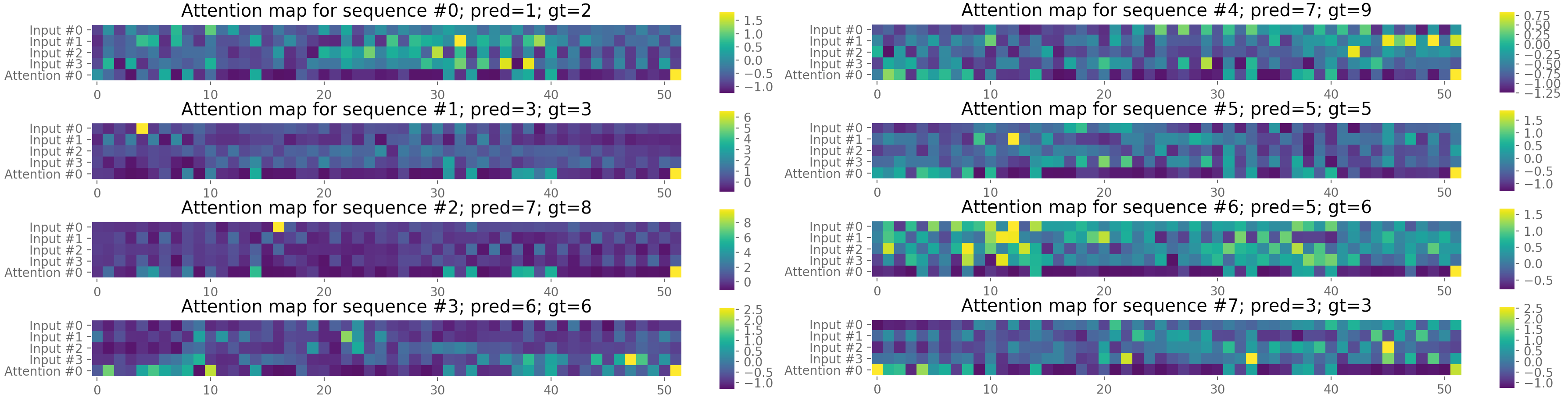}
    \vspace*{-0.5cm}
  \end{center}
  \caption{
          Attention distribution on selected events (best viewed in color)}
  \label{fig:ffa_attention}
\end{figure}

The results do present a bump in performance, with errors being distributed as $\sim \mathcal{N}(\mu=-0.081,\,\sigma^{2}=1.3)$. This means the feedforward attention mechanism provides a less biased estimator, with slightly less spread as well. The error distribution does still deviate for extreme events, where the standard deviation of the error becomes much higher. This reduces the confidence in our predictor that events that are predicted to have a high number of PVs actually do (hence reducing the confidence in our cut). The results and error distributions are shown in Fig.~\ref{fig:ffa_preds}.\\

Naturally, this also reduces the confidence in the attention maps, shown for a selected set of events in Fig.~\ref{fig:ffa_attention}. In such a high-dimensional space, and with little domain knowledge it is difficult to take advantage of the interpretability benefits of the attention mechanism. \\

There are nevertheless some potentially interesting hints to investigate further. For example, modules where a single sensor has a very high readout are often discarded by the attention mechanism, which would indicate that it roots out outliers such as those hypothesized in Section~\ref{subsec:regression}. We also see several modules being regularly attended to, such as modules \texttt{\#10} or \texttt{\#14} (which are very close to the beamspot), or modules at the extremes of the \velo detector. Additionally, we notice some decent recognition of track-like patterns, where a sequence of every other module is being fired (this is due to the ordering of the modules, with even-indexed and odd-indexed modules being split along the x-axis, see Fig.~\ref{fig:velo} (a)).

% Notes
% might be difficult to distinguish two close together PVs

\newpage
\section{Discussion}
\label{sec:discussion}

\subsection{Class imbalance and mapping complexity}
\label{subsec:imbalance}

Machine learning algorithms are data-hungry. An ever successful direction in the field has been to train more powerful models thanks to more data. Unfortunately, gathering such large amounts of data is time-consuming and painstaking. Nevertheless, our proposed approaches would certainly benefit from more training data, which, luckily, is readily available thanks to Monte-Carlo simulation. This is especially true in our context, where the input data is a compressed version of actual phenomena, with the raw bank sizes capturing the overall activity of a detector. Since many factors, with a large stochastic component, contribute to the activity to begin with, compressing this information makes it harder to ``extract''. This means it requires either a lot more data or a finer granularity to learn a precise mapping to the desired output, or a more explicit output.

This can particularly be an issue when looking to isolate the influence of some particular factor, as its influence can be drowned in that of many others, for example when trying to flag the presence or absence of a \B, as in Section~\ref{ssubsec:btag}. What renders this flagging even more difficult is the fact that proton-proton collisions are rich with physics: a given event will generate thousands of particles. \B-mesons will be present among these particles only in about 1 event in 20 (around 5\%), with often only one or two instances being generated. This means the signal we are searching for is buried very deep within the target, so we can expect a difficult learning process if we are expected to flag events based on the presence of \B-mesons.

While we propose certain methods to attempt to mitigate this class imbalance, the results seem to indicate that this is effectively a very difficult problem to solve. It shows that if there were to exist an explicit mapping between the compressed detector information and, for example, the presence of \B in an event, we would need much more data to pick up on the relevant discriminant patterns.

\subsection{Of the interpretability of neural networks}
\label{subsec:interpret}

One of the main advantages of algorithms that compute explicit transformations of their input (such as nearest neighbor algorithms or, more commonly in our setting, decision tree like random forests) is that they are said to be interpretable~\cite{Freitas:2014:CCM:2594473.2594475}. This means that the algorithms offer an explicit view of the decision process that produces a given output. The most common way of providing this feedback is to give the importance of input features. Note that this is very different from providing an \textit{explainable} model, which would show causal relationships, and could separate the explanation from the input mode~\cite{Lipton16a}. This setting is still very useful for the end-user, especially in domains such as physics where observations correlations between input and output can be directly highlighted. 

Attention mechanisms propose a straightforward solution to the long-term dependency problem in sequence modeling, by referring back to the input itself instead of attempting to build a local context as a hidden state. This same mechanism is also extremely useful for interpretability: by referring back to the input, we get a direct view of what the algorithm was considering as important to make a decision since this attention is learned through gradient descent jointly with the global objective.\\

This allows us to extend the set of tools that are usable in domains where interpretation is indeed very important, such as physics. By proposing a feed-forward alternative of the attention mechanism, we get a view into the decision process of a neural network as we would in a random forest, a popular algorithm for multivariate analysis, with sometimes the ability to have even richer feedback~\cite{olah2018the}.

\subsection{Future work}
\label{subsec:future}
The methods proposed here serve as a proof of concept, showing that data-driven methods could indeed be useful in a variety of ways. One of the most notable observations is that end-to-end solutions seem, as of now, rather limited in their possible applications. On the other hand decomposing the problem into smaller, learnable objectives looks like a promising direction.\\

For these methods to actually find their place in the analysis pipeline or in production software at \lhcb, a better characterization of their effect will be required. \\

Firstly, their computational overhead must be carefully considered. As is the case with most data-driven approaches, training complexity and stability is the main issue, whereas inference usually has constant-time complexity (or linear with the number of samples to be evaluated). A careful study of the input data, done to make sure the learned operational region is optimal for a given classifier may be needed. Ideally, one would prefer to leave the training data ``pure'', so as to build a truly end-to-end system. Realistically though, some outliers can prejudice the stability of the training process, hence the overall accuracy of a given predictor. Complexity-wise, the main constraint in production would be found at inference time: we want to introduce as little overhead as possible. As mentioned for example in Section~\ref{subsec:ffa}, feed-forward architectures are well suited to answer this problem. The One-Class SVM architecture proposed in Section~\ref{ssubsec:anomaly} also has a similar structure, with training being relatively costly - mainly because of the quadratic linear problem to be solved in finding the support vectors - and inference being linear on the number of support vectors.\\

Secondly, more work must be invested in characterizing the bias of these cuts. As has been discussed throughout this work, a predictor which consistently favors certain types of inputs is undesirable for production use if its effects cannot be corrected statistically. For example, we have shown in Section~\ref{ssubsec:anomaly} that the anomaly detection setting allows for decent prediction of the presence of \B-decays in a given event. It is still necessary to show that all \B decays are being treated equally by our predictor: if certain types of decays are favored, for example decays with muons compared to those without, or decays with shorter flight distances, then the filtered sample will be biased, which causes obvious issues for downstream analysis. If this was to be the case, one solution would be to implement finer-grained classifiers, each tuned to recognize certain decays. Their predictions could then be boosted or ensembled for a final prediction that could compensate some of these biases.

\newpage

\section{Conclusion}
\label{sec:conclusion}

The future upgrade of the \lhcb detector will require new approaches to handle an ever-increasing data rate. To do so in a real-time setting is an extremely challenging problem. In this work, we present some ideas for data-driven, physics-agnostic approaches, which present the advantage of being costly offline (training the algorithm to have the desired behavior) but cheap to evaluate online (constant inference time complexity).\\

We devise an empirical study of the feasibility of these data-driven approaches, notably their ability to learn or approximate the complex mappings between raw representations of a detector activity and physical quantities that could potentially be used downstream, for example in global event cuts.\\

Finally, we show that interpretability is not necessarily sacrificed when using these algorithms. To this end, we study a novel feed-forward attention mechanism, which points to important inputs as a way to interpret how a result was obtained. We release a library, implementing this algorithm in modern PyTorch~\cite{paszke2017automatic}, which can be used in a wide array of problems that could benefit from interpretability and/or relaxed sequence modeling techniques.\\

Overall these methods are incomplete to achieve the ideal end-to-end scenario, which likely requires much more advanced modeling and much more data. While the data-driven, heuristic-blind method is appealing at first glance, our results seem to indicate that domain knowledge is often still necessary. Including known priors, interpreting the results through a physical lens or choosing more manageable sub-problems thanks to domain knowledge are some examples of ways in which the expert is by no means replaceable by data.
\newpage

%\input{supplementary}

% Do not include this in any draft (just for information in the template)
%\input{acknowledgements_intro}

% $Id: appendix.tex 121212 2018-06-14 10:41:39Z pkoppenb $
% ===============================================================================
% Purpose: appendix to the standard template: standard symbol alises from Ulrik
% Author: Tomasz Skwarnicki
% Created on: 2009-09-24
% ===============================================================================

\clearpage

{\noindent\normalfont\bfseries\Large Appendices}

\appendix

\section{Visualization of the feature space}
\label{apx:manifold}

\begin{figure}[h]
  \begin{center}
    \includegraphics[width=1.0\linewidth]{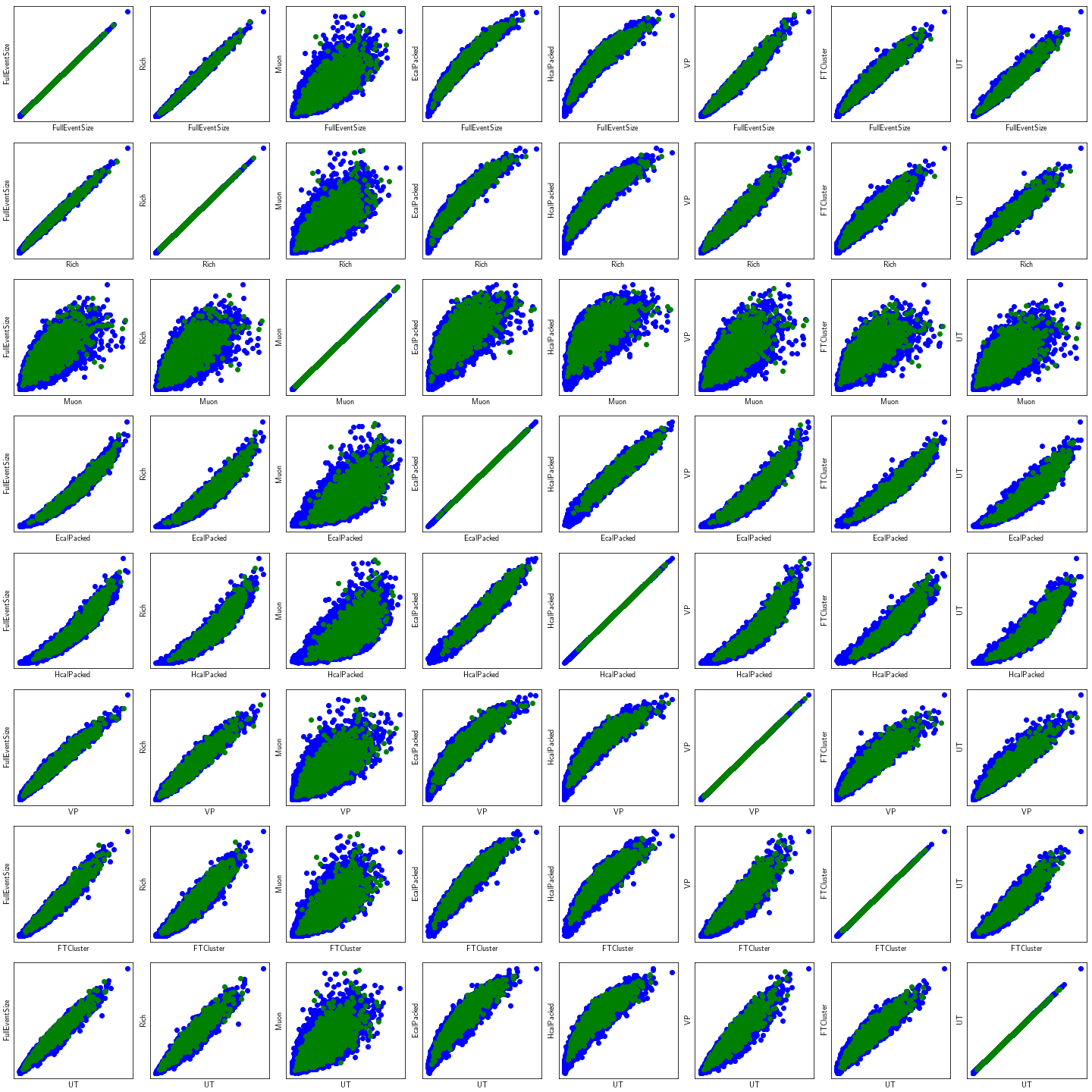}
    \vspace*{-0.5cm}
  \end{center}
  \caption{
  		Visualizing the feature space by projecting the data onto pairs of dimensions} 
  \label{fig:manifold}
\end{figure}

In Fig.~\ref{fig:manifold} we show the manifold formed by the data in the feature space, split between events containing at least a \B (in green) and those that don't (in blue). One interesting observation is that the events that contain \B lie close to the center of the manifold. This is expected, but reassuring: since the sensors are designed to detect this type of particle, the center of the manifold should be the operating region.

\section{Reconstructed vs Generated Primary Vertices}
\label{apx:genrec}

The Primary Vertex finding algorithms run in the \velo~\cite{Kucharczyk:1756296} do not accomplish a perfect reconstruction. This can be due to many factors, from discarding collisions that did not generate enough tracks to merging two tracks pertaining to two different but very close PVs.

\begin{figure}[h]
  \begin{center}
    \includegraphics[width=1.0\linewidth]{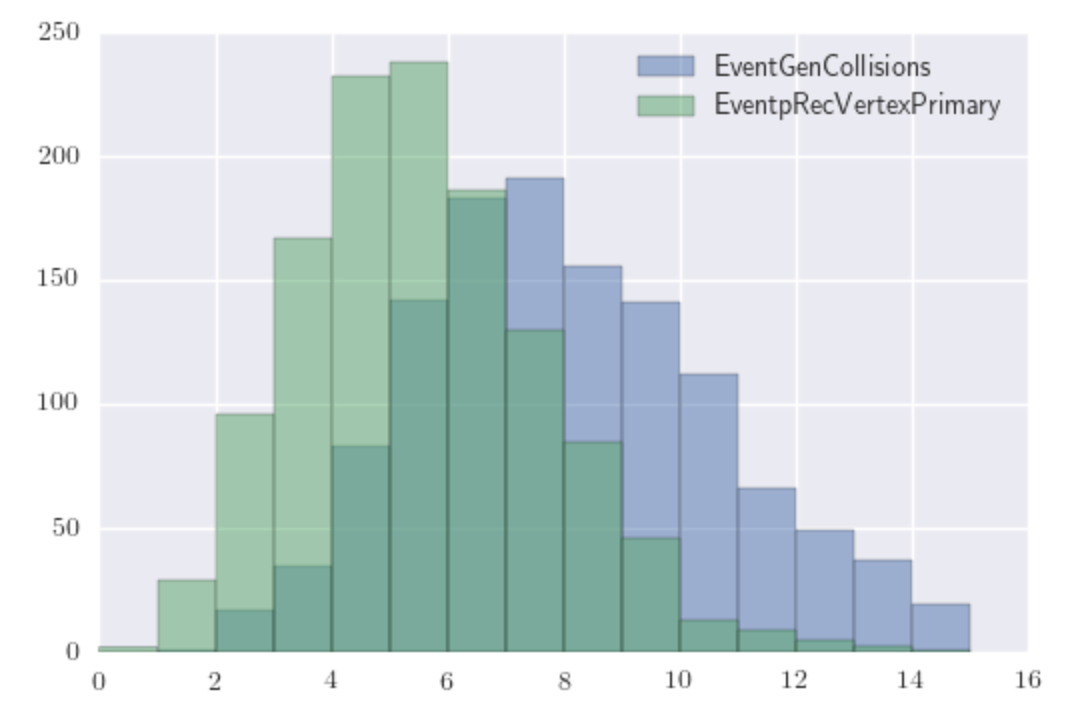}
    \vspace*{-0.5cm}
  \end{center}
  \caption{Distributions of Generated vs Reconstructed Primary Vertices} 
  \label{fig:recogen}
\end{figure}

In Fig.~\ref{fig:recogen}, we show both distributions. We observe a noticeable shift in the average number of Primary Vertices: on average 2 Primary Vertices are not reconstructed from the generated sample. We also note that it is possible for no Primary Vertices to be reconstructed, while it is impossible to be generated. 

We attempted most of our experiments on both target values. While they do not show any statistically significant benefit in terms of accuracy, this shows the advantage of data-driven methods, where the target can be easily swapped to any desirable correlated value.

\newpage

\section{Influence of $\nu$ in One-Class SVMs}
\label{apx:nusvm}

In this appendix section, we offer an additional visualization of the influence of the $\nu$ parameter in One-Class SVMs. This is the parameter that we control in Section~\ref{subsec:btag} to balance between throughput gain and signal loss. The discussion already include explanations of the parameters influence, but some additional figures are provided here to provide more intuition to the curious reader. 

\begin{figure}[h]
  \begin{center}
    \includegraphics[width=1.0\linewidth]{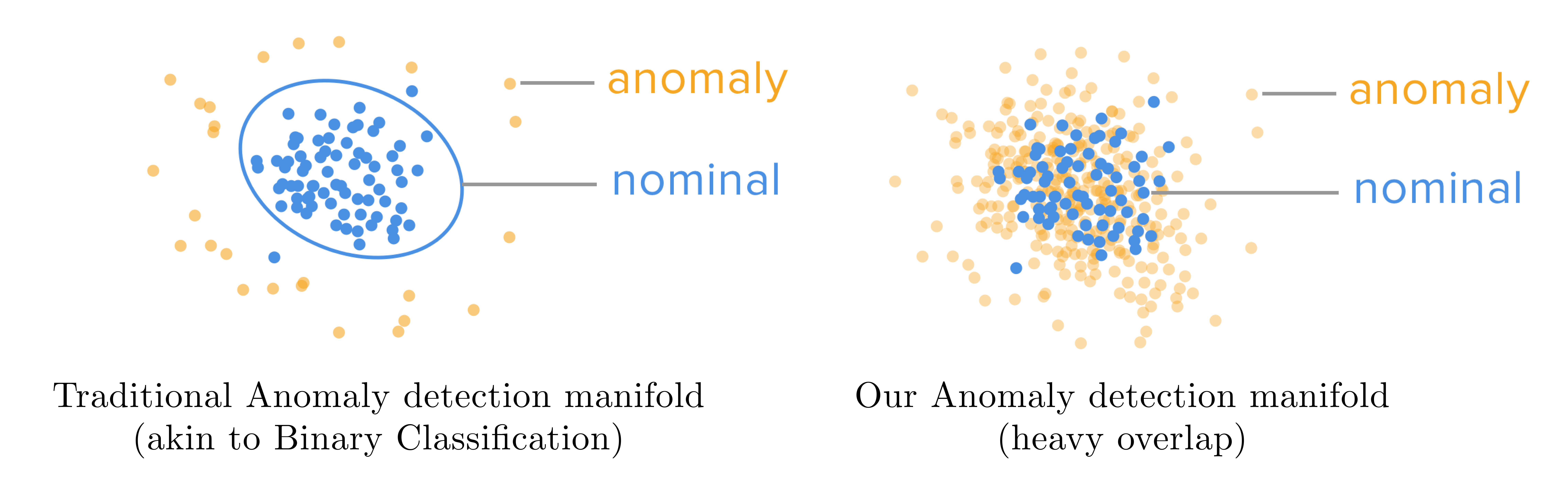}
    \vspace*{-0.5cm}
  \end{center}
  \caption{Comparison of ``classical'' (left) and our (right) settings of anomaly detection} 
  \label{fig:anom}
\end{figure}

\begin{figure}[h]
  \begin{center}
    \includegraphics[width=1.0\linewidth]{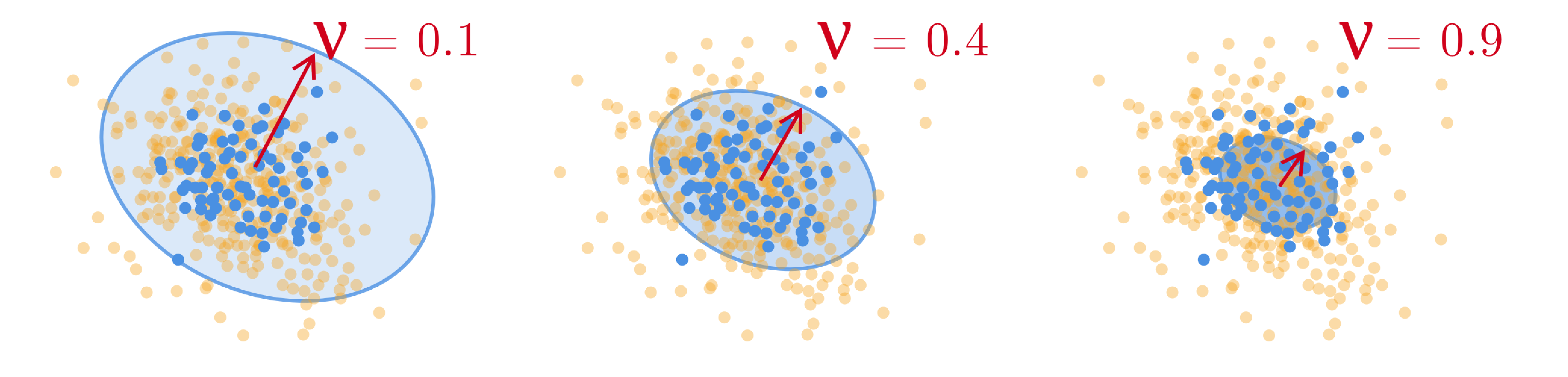}
    \vspace*{-0.5cm}
  \end{center}
  \caption{Influence of $\nu$ on the nominal class boundary} 
  \label{fig:nu_infl}
\end{figure}

Fig.~\ref{fig:anom} and Fig.~\ref{fig:nu_infl} illustrate with toy data what we mean by overlapping manifolds and how it motivates the choice of a parametric method.

\newpage

% This should be taken out in the final paper
%\input{supplementary-app}
\section*{Acknowledgements}
%
% These Acknowledgements valid from 20-Mar-2018
%
\noindent I would like to express gratitude towards the entire \lhcb collaboration, and particularly to the Trigger group for welcoming me to the team, to Conor Fitzpatrick and Sascha Stahl for their supervision and mentoring, Christoph Hasse for his high-quality tech support and overall support in this work, and Sepp Hochreiter's AI Lab in Linz for their help answering questions during their workshop. 

\addcontentsline{toc}{section}{References}
\setboolean{inbibliography}{true}
\bibliographystyle{LHCb}
%\bibliography{biblio,main,LHCb-PAPER,LHCb-CONF,LHCb-DP,LHCb-TDR}
\ifx\mcitethebibliography\mciteundefinedmacro
\PackageError{LHCb.bst}{mciteplus.sty has not been loaded}
{This bibstyle requires the use of the mciteplus package.}\fi
\providecommand{\href}[2]{#2}

\newpage

% Author List ----------------------------                                                                                                                                                                                                                                                                                                
%  You need to get a new author list!                                                                                                                                                                                                                                                                                                    

%\input{LHCb_HD_authorlist_2014-06-20}
 
\newpage
% \input{LHCb_authorlist.tex}
%
% The author list for journal publications is generated from the
% Membership Database shortly after 'approval to go to paper' has been
% given.  It will be sent to you by email shortly after a paper number
% has been assigned.  The author list should be included in the draft used for
% first and second circulation, to allow new members of the collaboration to verify
% that they have been included correctly. Occasionally a misspelled
% name is corrected, or associated institutions become full members.
% Therefore an updated author list will be sent to you after the final
% EB review of the paper.  In case line numbering doesn't work well
% after including the authorlist, try moving the \verb!\bigskip! after
% the last author to a separate line.
%
%
% The authorship for Conference Reports should be ``The LHCb
% collaboration'', with a footnote giving the name(s) of the contact
% author(s), but without the full list of collaboration names.

\end{document}